\documentclass[twocolumn, times]{aastex62}

\usepackage[caption=false]{subfig}

\usepackage{float,graphicx,amsmath,tabularx,booktabs,natbib,threeparttable}

\definecolor{dark-red}{rgb}{0.4,0.15,0.15}
\definecolor{dark-blue}{rgb}{0.15,0.15,0.4}
\definecolor{medium-blue}{rgb}{0,0,0.5}
\hypersetup{
    colorlinks, linkcolor={dark-red},
    citecolor={dark-blue}, urlcolor={medium-blue}
}
\usepackage{CJK}
\usepackage[whole]{bxcjkjatype}
\usepackage{CJKutf8}

\newcommand{\beqa}{\begin{eqnarray}} 
\newcommand{\eeqa}{\end{eqnarray}}

\newcommand{\bsub}{\begin{subequations}}
\newcommand{\esub}{\end{subequations}}
\newcommand{\beal}{\begin{align}}
\newcommand{\ealn}{\end{align}}

\graphicspath{{./}{figures/}}

\def \caltech {{Division of Physics, Mathematics and Astronomy, 
California Institute of Technology, Pasadena, CA 91125, USA}}

\def \coo {{Caltech Optical Observatories, California Institute of Technology, Pasadena, CA 91125, USA}}

\def \su {{Department of Astronomy, The Oskar Klein Center, Stockholm University, AlbaNova, 10691 Stockholm, Sweden}}

\begin{document}

\title{\Large{A Systematic Study of Ia-CSM Supernovae from the ZTF Bright Transient Survey}}

\author[0000-0003-4531-1745]{Yashvi Sharma}
\affiliation{\caltech}

\author[0000-0003-1546-6615]{Jesper Sollerman}
\affiliation{\su} 

\author[0000-0002-4223-103X]{Christoffer Fremling}
\affiliation{\caltech} 

\author[0000-0001-5390-8563]{Shrinivas R. Kulkarni}
\affiliation{\caltech} 

\author[0000-0002-8989-0542]{Kishalay De}
\affiliation{MIT-Kavli Institute for Astrophysics and Space Research, Cambridge, MA 02139, USA}

\author[0000-0002-7996-8780]{Ido Irani}
\affiliation{Department of Particle Physics and Astrophysics, Weizmann Institute of Science, 234 Herzl St, 76100 Rehovot, Israel} 

\author[0000-0001-6797-1889]{Steve Schulze}
\affiliation{Department of Physics, The Oskar Klein Center, Stockholm University, AlbaNova, 10691 Stockholm, Sweden} 

\author[0000-0002-4667-6730]{Nora Linn Strotjohann}
\affiliation{Department of Particle Physics and Astrophysics, Weizmann Institute of Science, 234 Herzl St, 76100 Rehovot, Israel} 

\author[0000-0002-3653-5598]{Avishay Gal-Yam}
\affiliation{Department of Particle Physics and Astrophysics, Weizmann Institute of Science, 234 Herzl St, 76100 Rehovot, Israel} 

\author[0000-0002-9770-3508]{Kate Maguire}
\affiliation{School of Physics, Trinity College Dublin, the University of Dublin, College Green, Dublin, Ireland}

\author[0000-0001-8472-1996]{Daniel A. Perley}
\affiliation{Astrophysics Research Institute, Liverpool John Moores University, Liverpool Science Park, 146 Brownlow Hill, Liverpool L35RF, UK}

\author[0000-0001-8018-5348]{Eric C. Bellm}
\affiliation{DIRAC Institute, Department of Astronomy, University of Washington, 3910 15th Avenue NE, Seattle, WA 98195, USA}

\author[0000-0002-7252-3877]{Erik C. Kool}
\affiliation{\su}

\author[0000-0001-5955-2502]{Thomas Brink}
\affiliation{Department of Astronomy, University of California, Berkeley, CA 94720-3411, USA}

\author[0000-0001-8208-2473]{Rachel Bruch}
\affiliation{Department of Particle Physics and Astrophysics, Weizmann Institute of Science, 234 Herzl St, 76100 Rehovot, Israel}

\author[0000-0001-8857-9843]{Maxime Deckers}
\affiliation{School of Physics, Trinity College Dublin, the University of Dublin, College Green, Dublin, Ireland}

\author[0000-0002-5884-7867]{Richard Dekany}
\affiliation{\coo}

\author[0000-0001-7344-0208]{Alison Dugas}
\affiliation{Institute for Astronomy, University of Hawai’i, Honolulu, HI 96822, USA}

\author{Samantha Goldwasser}
\affiliation{Department of Particle Physics and Astrophysics, Weizmann Institute of Science, 234 Herzl St, 76100 Rehovot, Israel} 

\author{Matthew J. Graham}
\affiliation{\caltech}

\author[0000-0002-9154-3136]{Melissa L. Graham}
\affiliation{DIRAC Institute, Department of Astronomy, University of Washington, 3910 15th Avenue NE, Seattle, WA 98195, USA}

\author[0000-0001-5668-3507]{Steven L. Groom}
\affiliation{IPAC, California Institute of Technology, 1200 E. California
             Blvd, Pasadena, CA 91125, USA}

\author[0000-0001-9315-8437]{Matt Hankins}
\affiliation{Arkansas Tech University, Russellville, AR 72801, USA}

\author[0000-0001-5754-4007]{Jacob Jencson}
\affiliation{Steward Observatory, University of Arizona, 933 North Cherry Avenue, Tucson, AZ 85721-0065, USA}

\author[0000-0001-5975-290X]{Joel P. Johansson}
\affiliation{Department of Physics, The Oskar Klein Center, Stockholm University, AlbaNova, 10691 Stockholm, Sweden} 

\author[0000-0003-2758-159X]{Viraj Karambelkar}
\affiliation{\caltech} 

\author[0000-0002-5619-4938]{Mansi M. Kasliwal}
\affiliation{\caltech}

\author[0000-0002-8532-9395]{Frank J. Masci}
\affiliation{IPAC, California Institute of Technology, 1200 E. California
             Blvd, Pasadena, CA 91125, USA}
             
\author[0000-0002-7226-0659]{Michael S. Medford}
\affiliation{Department of Astronomy, University of California, Berkeley, Berkeley, CA 94720}
\affiliation{Lawrence Berkeley National Laboratory, 1 Cyclotron Rd., Berkeley, CA 94720}

\author[0000-0002-0466-1119]{James D. Neill}
\affiliation{\caltech} 

\author[0000-0002-7501-5579]{Guy Nir}
\affiliation{Department of Astronomy, University of California, Berkeley, CA 94720-3411, USA}

\author[0000-0002-0387-370X]{Reed L. Riddle}
\affiliation{\coo}

\author{Mickael Rigault}
\affiliation{Université Clermont Auvergne, CNRS/IN2P3, Laboratoire de Physique de Clermont, 63000 Clermont-Ferrand, France}

\author[0000-0001-8948-3456]{Tassilo Schweyer}
\affiliation{\su}

\author[0000-0001-9834-3439]{Jacco H. Terwel}
\affiliation{School of Physics, Trinity College Dublin, the University of Dublin, College Green, Dublin, Ireland}
\affiliation{Isaac Newton Group (ING), Apt. de correos 321, E-38700, Santa Cruz de La Palma, Canary Islands, Spain}

\author[0000-0003-1710-9339]{Lin Yan}
\affiliation{\caltech} 

\author[0000-0002-6535-8500]{Yi Yang\begin{CJK*}{UTF8}{gbsn}
(杨轶)
\end{CJK*}}
\affiliation{Department of Astronomy, University of California, Berkeley, CA 94720-3411, USA}

\author[0000-0001-6747-8509]{Yuhan Yao}
\affiliation{\caltech} 

\correspondingauthor{Yashvi Sharma}
\email{yssharma@astro.caltech.edu}


\begin{abstract}
    Among the supernovae (SNe) that show strong interaction with the circumstellar medium, there is a rare subclass of Type Ia supernovae, SNe Ia-CSM, that show strong narrow hydrogen emission lines much like SNe IIn but on top of a diluted over-luminous Type Ia spectrum. In the only previous systematic study of this class \citep{Silverman_2013}, 16 objects were identified, 8 historic and 8 from the Palomar Transient Factory (PTF). Now using the successor survey to PTF, the Zwicky Transient Facility (ZTF), we have classified 12 additional objects of this type through the systematic Bright Transient Survey (BTS). In this study, we present and analyze the optical and mid-IR light curves, optical spectra and host galaxy properties of this sample. Consistent with previous studies, we find the objects to have slowly evolving light curves compared to normal SNe Ia with peak absolute magnitudes between $-19.1$ and $-21$, spectra having weak H$\beta$, large Balmer decrements of $\sim7$ and strong Ca NIR emission. Out of 10 SNe from our sample observed by NEOWISE, 9 have $3\sigma$ detections, along with some showing a clear reduction in red-wing of H$\alpha$, indicative of newly formed dust. We do not find our SN Ia-CSM sample to have significantly different distribution of equivalent width of \ion{He}{1} $\lambda5876$ than SNe IIn as observed in \citet{Silverman_2013}. The hosts tend to be late-type galaxies with recent star formation. We also derive a rate estimate of 29$^{+27}_{-21}$\,Gpc$^{-3}$\,yr$^{-1}$ for SNe Ia-CSM which is $\sim$0.02--0.2\% of the SN Ia rate. This work nearly doubles the sample of well studied Ia-CSM objects in \citet{Silverman_2013}, increasing the total number to 28.
\end{abstract}

\section{Introduction} 
When it comes to supernovae (SNe) interacting with circumstellar material (CSM), a number of sub-types of core-collapse SNe (CCSNe) show signs of strong interaction, 
like SNe~IIn \citep{Schlegal1990,Fillipenko1997}, SNe~Ibn \citep{Pastorello2008a,Foley_2007,Chugai2009,Hosseinzadeh_2017} and most recently SNe~Icn \citep{Icn2021,Galyam2022Icn,Perley2021Icn}. SN~IIn progenitors are generally thought to be massive stars (like Luminous Blue Variables, LBVs) that lose their hydrogen envelopes to wind-driven mass loss and outbursts \citep{Galyam2007,Galyam2009,Kiewe2012,Taddia2013,smith2014}. Helium-rich but hydrogen-deficient CSM in the case of SNe~Ibn \citep{Pastorello2008a,Foley_2007,Chugai2009} and both hydrogen and helium deficient CSM in SNe~Icn \citep{Galyam2022Icn,Perley2021Icn,Pellegrino2022} are thought to arise from high-velocity wind mass loss or stripping of the envelope in binary configurations of massive Wolf-Rayet (WR) like stars. For SNe IIn in most cases, the mass-loss rate derived from the CSM velocity is consistent with estimates from LBV-like eruptive mass loss.

However, there exists a rare sub-type of thermonuclear supernovae (SNe Ia) which also interacts strongly with CSM i.e. SNe Ia-CSM. This class poses a challenge to the progenitor debate of SNe Ia. There is some consensus on there being at least two major progenitor channels for SNe Ia; the double-degenerate (DD) channel \citep{Webbink1984,IbenTutukov1984} which is the merging of two C/O white dwarfs and the single-degenerate (SD) channel \citep{WhelanIben1973} where the white dwarf accretes enough material from a non-degenerate companion to explode. Although there are more arguments for the DD scenario from observations of nearby SNe Ia \citep{Nugent2011,Li2011,Brown_2012,Bloom_2011}, the strongest observational evidence for the SD scenario are SNe Ia with CSM. 

Indications of CSM around SNe Ia ranges from detection of time varying narrow \ion{Na}{1}D absorption lines \citep{Patat2007,Blondin_2009,Simon_2009} in high-resolution spectra (found in at least 20\% of SNe Ia in spiral hosts, \citealt{Sternberg2011,maguire2013,clark2021}), to strong intermediate and narrow Balmer emission features in the spectra and large deviations of the light curves from the standard shape. The latter phenomena have been named SNe Ia-CSM \citep{Silverman_2013}, but were earlier referred to as ``SNe IIna" or ``SNe Ian" due to the strong similarity between their spectra and those of  SNe IIn. The first two examples of this class studied in detail were SNe 2002ic \citep{Hamuy2003,Deng_2004,Wang_2004,Wood_Vasey_2004,Kotak2005,chugai2004} and 2005gj \citep{Aldering2006,Prieto2007}, but for a long time there was ambiguity regarding their thermonuclear nature \citep{Benetti2006}. These SNe were dominated by interaction from the first spectrum and were quite over-luminous compared to normal SNe Ia. The first clear example of a thermonuclear SN Ia-CSM was PTF11kx \citep{dilday2012,Silverman2013_11kx}. It looked like a luminous SN Ia (99aa-like) at early phases but started showing interaction at $\sim60$ days from explosion and thereafter strongly resembled SNe 2002ic and 2005gj at late times. Higher resolution spectra taken at early times indicated multiple shells of CSM with some evacuated regions in between. \citet{dilday2012} suggested a symbiotic nova progenitor involving a WD and a red giant (similar to RS Ophiuchi) could produce such CSM distribution, however later studies argued that the massive CSM of PTF11kx was inconsistent with the mass-loss rates from symbiotic nova systems \citep{Silverman2013_11kx, Soker2013}.

Ever since, a handful of SNe of this class have been studied in detail to investigate their progenitors and to distinguish them from their spectroscopic cousins, the Type IIn SNe. Both SN Ia-CSM and SN IIn spectra share a blue quasi-continuum, a strong H$\alpha$ feature with an intermediate and a narrow component, and often a broad Ca NIR triplet feature, but they differ with regards to the line strength of H$\beta$, strength/presence of helium and presence of emission lines from intermediate mass elements often found in CCSNe. There are some individual SNe with unclear type often referred to as SN Ia-CSM/IIn, like SN 2012ca for which some papers argue for core-collapse \citep{inserra2014,inserra2016} and others for a thermonuclear origin \citep{fox2015}. This ambiguity becomes more dominant as the underlying SN flux gets smaller compared to the interaction power \citep{Leloudas2015}. \citet[hereafter S13]{Silverman_2013} is the only study to analyze a sample of SNe Ia-CSM, 16 objects in total including 6 previously known, 3 re-discovered (re-classified SNe IIn) and 7 new from the Palomar Transient Factory (PTF). Their paper presents the common properties of optical light curves, spectra and host galaxies and contrast them against SN IIn properties. In this paper, we present 12 new SNe Ia-CSM discovered as part of the Zwicky Transient Facility's (ZTF; \citealp{Bellm2019b,graham2019,Dekany20}) Bright Transient Survey (BTS; \citealp{Fremling2020, Perley2020}) and analyze their optical light curves, spectra, hosts and rates. Throughout this paper, we have compared the results derived from our sample to the ones in S13.

This paper is organised as follows; we first discuss the sample selection criteria, the photometric and spectroscopic data collection in \S2, then the analysis of light- and color-curves and the bolometric luminosities is done in \S3.1. The analysis of early and late-time spectra and emission line identification is presented in \S3.2, and analysis of the host galaxies is provided in \S3.3. The rates are estimated from the BTS survey in \S3.4. We end with a discussion about the nature of SN Ia-CSM progenitors and a summary in \S4 and \S5.

\section{Observations and Data reduction}
In this section, we outline our selection criteria, and present the optical photometry and spectroscopic observations of the 12 SNe Ia-CSM in our sample.

\subsection{Selection Criteria}
To carefully curate our sample of SNe Ia-CSM, we used the BTS sample and its publicly available BTS Sample Explorer\footnote{\url{https://sites.astro.caltech.edu/ztf/bts/explorer.php}} website to obtain the list of all classified Type Ia sub-types during the period 2018-05-01 to 2021-05-01. We then filter out oddly behaving Type Ia SNe based on their light-curve properties. We used two criteria; the primary being rest-frame duration considering flux above 20\% of peak flux, and the second being change in magnitude after 30 days from peak ($\Delta m_{30}$). We calculated these two properties from either $g$ or $r$-band light curves (whichever had maximum number of detections) grouped into 3-day bins and used Gaussian Process Regression\footnote{\citet{scikit-learn} \url{https://scikit-learn.org/stable/modules/gaussian_process.html}} to interpolate the light curves where coverage was missing. For the first filtering, we calculated the mean ($\mu \approx 35$ days) and standard deviation ($\sigma \approx 16$ days) of the duration distribution and selected everything that had a duration greater than $\mu + 3\sigma$. Given the large sample size ($N=3486$), the standard error on the mean is $\sim 0.5$ days, hence our duration cut of $3\sigma$ is suitable. This filtering selected 41 out of 3486 BTS SNe Ia. Then from these 41 SNe, we calculated the mean and standard deviation of the $\Delta m_{30}$ distribution and removed SNe that were more than 1$\sigma$ away from the mean on the higher side to reject the relatively steeply declining long SNe, which resulted in 35 SNe being kept. Again, the mean and standard deviation of $\Delta m_{30}$ distribution of these 41 long duration SNe are 0.48 mag and 0.27 mag respectively and the standard error on mean is $\sim0.04$, making our $1\sigma$ cut suitable. Finally, we manually inspected the 35 selected SNe Ia to confirm their classification. 20 out of the 35 SNe that passed the above filtering criteria were just normal SNe Ia either caught late or missing some post-peak coverage in ZTF or had spurious detections that resulted in long duration estimates, 2 had incorrect duration estimate due to an interpolation error and were recalculated and 1 (AT2020caa; \citealp{ZTF20aamibse}) had some detections before the SN explosion which could be connected to a different SN (i.e. a sibling; \citealp{Graham2022}).

The remaining 12 long-duration SNe Ia all turned out to be spectroscopically classified SNe Ia-CSM in BTS, and none of the classified BTS SNe Ia-CSM were missed in this filtering. No other SNe apart from these stood out in particular, indicating the classification reliability of the BTS sample. During the same period, 9 SNe Ia-CSM were reported to the Transient Name Server (TNS), out of which 7 are already in our sample, 1 was detected by ZTF but did not meet the BTS criteria, and 1 was not detected in ZTF as the transient location fell too close to the field edges and was masked by the automated image subtraction pipeline. \citet{Yao2019} presented early photometric observations of one SN Ia-CSM in our sample, SN 2018crl. Table~\ref{tab:prop} summarizes the coordinates, redshifts, peak absolute magnitudes, durations, host galaxy information and Milky Way extinction for the 12 SNe Ia-CSM in our sample.

Furthermore, we re-checked the classifications of 142 SNe IIn classified in BTS during the same period as above, in case any SN Ia-CSM was masquerading among them and found 6 to have ambiguous classifications. These are discussed further in Appendix~\ref{app:A}.

\begin{table*}[!t]
    \centering
    \caption{Properties of the 12 BTS SNe Ia-CSM}
    \begin{threeparttable}
    \begin{tabularx}{0.85\textwidth}{ccccccc}
    \toprule
    \toprule
        \textbf{ZTF Name} & \textbf{IAU Name} & \textbf{z} & \textbf{$M_{r}^{\rm{peak}}$} & \textbf{Duration\tnote{1}} & \textbf{Host Name} & \textbf{Host Mag\tnote{2}} \\
         & & & (mag) & \text{(days)} & & ($m_{r}$) \\
        \midrule
        ZTF18aaykjei & SN 2018crl & 0.097 & -19.66 & 130 & SDSS J161938.90+491104.5 & 18.89 \\
        ZTF18abuatfp & SN 2018gkx & 0.1366 & -20.07 & 322 & SDSS J135219.22+553830.2 & 18.23 \\
        ZTF18actuhrs & SN 2018evt & 0.02378 & -19.10 & 447 & MCG-01-35-011 & 14.07 \\
        ZTF19aaeoqst & SN 2019agi & 0.0594 & $<$-18.76 & $>$303 & SDSS J162244.06+240113.4 & 17.82 \\
        ZTF19abidbqp & SN 2019ibk & 0.04016 & $<$-17.55 & $>$576 & SDSS J014611.93-161701.1 & 15.55 \\
        ZTF19acbjddp & SN 2019rvb & 0.1835 & -20.74 & 172 & WISEA J163809.90+682746.3 & 20.44 \\
        ZTF20abmlxrx & SN 2020onv & 0.095 & $<$-20.36 & $>$154 & WISEA J231646.31-231839.9 & 17.95 \\
        ZTF20abqkbfx & SN 2020qxz & 0.0964 & -20.00 & 166 & WISEA J180400.99+740050.0 & 17.65 \\
        ZTF20accmutv & SN 2020uem & 0.041 & $<$-20.17 & $>$279 & WISEA J082423.32-032918.6 & 15.88 \\
        ZTF20aciwcuz & SN 2020xtg & 0.0612 & $<$-19.60 & $>$336 & SDSS J153317.64+450022.8 & 15.42 \\
        ZTF20acqikeh & SN 2020abfe & 0.093 & -20.24 & 171 & SDSS J200003.30+100904.2 & 20.18 \\
        ZTF21aaabwzx & SN 2020aekp & 0.046 & -19.62 & 458 & SDSS J154311.45+174843.7 & 18.41 \\ 
        \bottomrule
    \end{tabularx}
    \begin{tablenotes}
    \item[1] Rest frame duration above 20\% of $r$-band peak flux, uncertainty of $\pm2-3$\,days from ZTF cadence.
    \item[2] Corrected for Galactic extinction.
    \end{tablenotes}
    \end{threeparttable}
    \label{tab:prop}
\end{table*}

\subsection{Discovery}
All SNe Ia-CSM were detected by the ZTF \citep{Bellm2019b,graham2019,Dekany20} and passed the criteria for the BTS \citep{Fremling2020, Perley2020} automatic filtering, i.e.\ extra-galactic real transients with peak magnitudes brighter than 19 mag. These were saved and classified as part of BTS which aims to classify all transients brighter than 18.5 magnitude, and reported to the Transient Name Server\footnote{https://www.wis-tns.org/} (TNS) during the period 2018-05-01 to 2021-05-01. Out of the 12 SNe, 6 were first reported to TNS (i.e.\ discovered) by ZTF (AMPEL, \citealt{ampel2019,catsHTM2018} and BTS), 3 were first reported by GaiaAlerts \citep{gaia2021}, 2 by ATLAS \citep{atlassmith2020} and 1 by ASAS-SN \citep{asassn2014}. For classification, 9 were classified by the ZTF group, 1 by ePESSTO \citep{epessto2015,TNS2018evt}, 1 by SCAT \citep{scat2018,TNS2019agi} and 1 by the Trinity College Dublin (TCD) group \citep{TNS2020onv}. The follow-up spectral series for these SNe were obtained as part of the BTS classification campaign as many were difficult to classify with the ultra-low resolution spectrograph P60/SEDM \citep{sedm2018} and hence were followed up with intermediate resolution spectrographs. The SEDM spectra were helpful in determining an initial redshift but the template matches were unclear (matched to SN IIn as well as SN Ia-CSM and SN Ia-pec templates, some matched poorly to SN Ia/Ic at early times). SNe 2019agi (classification and spectrum taken from TNS), 2019rvb, 2020onv, 2020qxz and 2020uem were classified as Ia-CSM $\sim1-2$ month after discovery using spectra at phases of 42, 26, 38, 45 and 51 days respectively. SNe 2018crl, 2018gkx and 2019ibk were classified $\sim2-3$ months after discovery using spectra at phases of 92, 75 and 103 days respectively. SNe 2018evt, 2020abfe and 2020aekp were classified $\sim4-5$ months after discovery using the spectra at phases of 144, 146 and 132 days respectively. SN 2020xtg immediately went behind the sun after its first detection in ZTF therefore its first spectrum (using SEDM) was taken at 91 days since explosion which was dominated by strong H$\alpha$ emission, and thus SN 2020xtg was initially classified as a Type II. As this SN was exhibiting a long lasting light curve, an intermediate resolution spectrum was taken at 340 days which matched very well to SN Ia-CSM and therefore its classification was updated. SNe 2020uem and 2020aekp showed peculiar features and were followed up for more optical spectroscopy for single object studies (to be presented in future papers). 

\subsection{Optical photometry}\label{sec:dataphot}

To assemble our sample light curves, we obtained forced PSF photometry via the ZTF forced-photometry service \citep{Masci2019,ztfipac_doi} in $g$, $r$ and $i$ bands and also added data from ATLAS \citep{Tonry2018,atlassmith2020} forced-photometry service in $c$ and $o$ bands. The high cadence ZTF partnership survey in $i$ band contributed some photometry to SNe 2018crl, 2018gkx, 2019agi, 2019ibk and 2019rvb. The ZTF and ATLAS data were supplemented with data from the Rainbow camera (RC, \citealt{Ben-Ami12}) on the robotic Palomar 60-inch telescope \citep[P60,][]{Cenko2006} and the Optical wide field camera (IO:O) on the Liverpool telescope \citep[LT,][]{lt}. The P60 data was processed with the automatic image subtraction pipeline \texttt{FPipe} \citep{Fremling2016} using reference images from SDSS when available, and otherwise from Pan-STARRS1. The IO:O data was initially reduced with their standard pipeline\footnote{\url{https://telescope.livjm.ac.uk/TelInst/Pipelines/}} then image subtraction was carried out using the method outlined in \citet{ljmu14104}. For SN 2018evt, some early time data available from ASAS-SN \citep{asassn2014,asassn2017} in the $V$ band was obtained through their \textit{Sky Patrol}\footnote{https://asas-sn.osu.edu/} interface.

\begin{figure*}[!h]
    \centering
    \includegraphics[width=\textwidth]{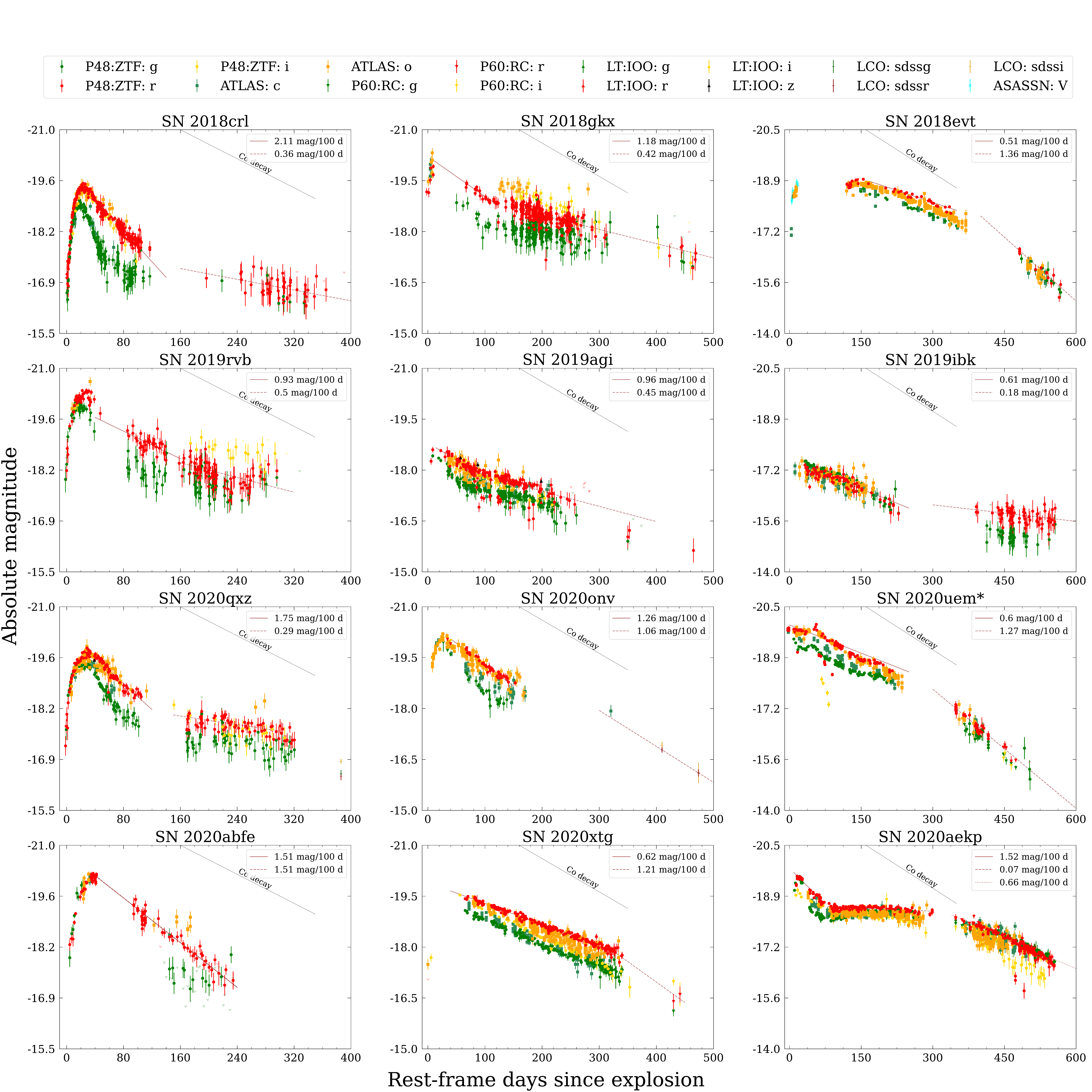}
    \caption{Optical light curves of the ZTF BTS SN Ia-CSM sample. The SNe Ia-CSM have longer duration than the average SN Ia, with some variety like bumpy light curves or long plateaus. The one SN marked with an asterisk (SN 2020uem) has an unconstrained explosion time estimate ($\sim\pm50$\,d). The decline rate from Cobalt decay is marked with black dashed line, the light curve decline rates measured from $r$-band data are shown in the subplot legends.}
    \label{fig:lc_collage}
\end{figure*}

We corrected all photometry for Milky Way extinction with the Python package \texttt{extinction} \citep{extinction2016} using the dust extinction function from \citet{fitzp1999}, the \citet{Schlafly_2011} dust map, and an R$_V$ of 3.1. Then we converted all measurements into flux units for analysis and considered anything less than a 3$\sigma$ detection an upper limit. There is moderate to good coverage in $g$, $r$, $c$ and $o$ bands for all SNe in our sample. Figure~\ref{fig:lc_collage} shows a multi-paneled figure of the light curves of the objects in our sample. 

\subsection{Mid-IR photometry}
The transients were observed during the ongoing NEOWISE all-sky mid-IR survey in the $W1$ ($3.4$\,$\mu$m) and $W2$ ($4.5$\,$\mu$m) bands \citep{Wright2010, Mainzer2014}. We retrieved time-resolved coadded images of the field created as part of the unWISE project \citep{Lang2014, Meisner2018}. To remove contamination from the host galaxies, we used a custom code \citep{De2019} based on the ZOGY algorithm \citep{Zackay2016} to perform image subtraction on the NEOWISE images using the full-depth coadds of the WISE and NEOWISE mission (obtained during 2010-2014) as reference images. Photometric measurements were obtained by performing forced PSF photometry at the transient position on the subtracted WISE images until the epoch of the last NEOWISE data release (data acquired until December 2021). Further analysis of the mid-IR photometry is presented in \S\ref{midir}

\subsection{Optical spectroscopy}\label{sec:dataspec}
 The main instruments used for taking spectra and the software used to reduce the data are summarized in Table~\ref{tab:inst}. Additionally, the spectrum \citet{afosch_spec4_cmutv} obtained using the Asiago Faint Object Spectrograph and Camera (AFOSC) on the 1.8~m telescope at Cima Ekar, and the spectrum \citet{ntt_spec4_tuhrs} obtained using the ESO Faint Object Spectrograph and Camera version 2 (EFOSC2) on ESO New Technology Telescope (NTT) were taken from TNS. 
 
\begin{table}[]
    \centering
    \caption{Description of spectrographs used for follow-up and the corresponding data reduction pipelines}
    \small
    \begin{threeparttable}
    \begin{tabularx}{0.48\textwidth}{ccc}
    \toprule
    \toprule
        \textbf{Inst.} & \textbf{Telescope} & \textbf{Reduction Software} \\
        \midrule
        SEDM\tnote{1} & Palomar 60-inch (P60) & \texttt{pySEDM}\tnote{2} \\
        ALFOSC\tnote{3} & Nordic Optical Telescope & IRAF\tnote{4}, PyNOT\tnote{14}, \texttt{pypeit} \\
        DBSP\tnote{5} & Palomar 200-inch (P200) & IRAF\tnote{6}, \texttt{DBSP\_DRP}\tnote{7} \\
        KAST\tnote{8} & Shane 3-m & IRAF \\
        LRIS\tnote{9} & Keck-I & LPipe\tnote{10} \\
        SPRAT\tnote{11} & Liverpool Telescope & \citet{frodospec2012} \\
        DIS\tnote{12} & APO\tnote{13} & IRAF \\
        
        \bottomrule
    \end{tabularx}
    \begin{tablenotes}
    \item[1] Spectral Energy Distribution Machine \citep{sedm2018}
    \item[2] \citet{pysedm2019}
    \item[3] Andalucia Faint Object Spectrograph and Camera
    \item[4] \citet{Tody1986, Tody1993}
    \item[5] Double Beam Spectrograph \citep{Oke1982}
    \item[6] Standard pipeline by \citet{Bellm2016} used prior to Fall 2020
    \item[7] \texttt{pypeit} \citep{pypeit:joss_pub} based pipeline (\url{https://github.com/finagle29/dbsp_drp}) used since Fall 2020
    \item[8] Kast Double Spectrograph \citep{kast}
    \item[9] Low Resolution Imaging Spectrometer \citep{Oke1995}
    \item[10] IDL based automatic reduction pipeline\footnote{\url{https://sites.astro.caltech.edu/~dperley/programs/lpipe.html}} \citep{Perley2019}
    \item[11] Spectrograph for the Rapid Acquisition of Transients \citep{SPRAT}
    \item[12] Dual Imaging Spectrograph
    \item[13] Astrophysics Research Consortium telescope at the Apache Point Observatory
    \item[14] \url{https://github.com/jkrogager/PyNOT}
    \end{tablenotes}
    \end{threeparttable}
    \label{tab:inst}
    \hrule
\end{table}

The details for all optical spectra (61 for the sample in total) presented in this paper are provided in Table~\ref{tab:samplespec}. Furthermore, all spectra were corrected for Milky Way extinction using
\texttt{extinction} and the same procedure as for the photometry.
The SN redshifts were derived using narrow host lines for the objects which did not already have a host redshift available in the NASA/IPAC Extragalactic Database\footnote{https://ned.ipac.caltech.edu/} (NED). Photometric calibration was done for all spectra i.e. they were scaled such that the synthetic photometry from the spectrum matched the contemporaneous host-subtracted ZTF $r$-band data. For SN 2018crl, a host galaxy spectrum taken using P200/DBSP was available, which was subtracted from the P200/DBSP SN spectrum taken at +92 days. For SN 2020aekp, more spectra beyond $\sim350$ days were obtained but will be presented in a future study of the object (34 additional spectra up to $\sim$600 day).

These processed spectra were used for the rest of the analysis as detailed in $\S\ref{sec:specanalysis}$ and will be available on WISeREP\footnote{https://www.wiserep.org/} \citep{wiserep}.

\begin{table*}[]
    \centering
    \caption{Summary of optical spectra}
    \footnotesize
    \begin{tabularx}{0.95\textwidth}{c c c c c || c c c c c}
    \toprule
    \toprule
\textbf{SN} & \textbf{JD} & \textbf{Epoch} & \textbf{Telescope/Instrument} & \textbf{Int} & \textbf{SN} & \textbf{JD} & \textbf{Epoch} & \textbf{Tel./Instr.} & \textbf{Int} \\
 & ($-2450000$) & (days) &  & (sec) & & ($-2450000$) & (days) &  & (sec) \\
 \midrule
SN 2018crl & 8282 & 9 & APO/DIS & 2400 & SN 2020uem & 9128 & 11 & P60/SEDM & 1800 \\
 & 8288 & 15 & P60/SEDM & 2700  & & 9136 & 18 & P60/SEDM & 1800 \\                 
 & 8295 & 21 & P60/SEDM & 2700  & & 9170 & 51 & Ekar/AFOSC & 1200 \\
 & 8306 & 31 & P60/SEDM & 2700  & & 9222 & 101 & Lick-3m/KAST & 3600 \\
 & 8373 & 92 & P200/DBSP & 600 & & 9252 & 130 & Lick-3m/KAST & 2700\\
(Host) & 8627 & 324 & P200/DBSP & 900 & & 9263 & 140 & Lick-3m/KAST & 2400\\
SN 2018gkx & 8457 & 75 & Keck1/LRIS & 300 & & 9291 & 167 & NOT/ALFOSC & 900 \\
SN 2018evt & 8343 & 9 & NTT/EFOSC2 & 300 & & 9481 & 349 & P60/SEDM & 2160 \\
 & 8465 & 127 & P60/SEDM & 1200 & & 9492 & 360 & Keck1/LRIS & 600 \\
 & 8481 & 143 & P60/SEDM & 1200 & & 9583 & 448 & P60/SEDM & 2160 \\
 & 8481 & 144 & LT/SPRAT & 1000 & & 9586 & 451 & P60/SEDM & 2160 \\
 & 8534 & 195 & P60/SEDM & 1200 & SN 2020xtg & 9226 & 91 & P60/SEDM & 2160 \\
SN 2019agi & 8547 & 42 & UH88/SNIFS & 1820 & & 9491 & 340 & Keck1/LRIS & 600 \\
SN 2019ibk & 8691 & 35 & P60/SEDM & 2250 & & 9606 & 448 & Keck1/LRIS & 1200 \\
 & 8695 & 39 & P60/SEDM & 2250 & SN 2020abfe & 9189 & 27 & P60/SEDM & 2700 \\
 & 8697 & 41 & P60/SEDM & 2250 & & 9319 & 146 & Keck1/LRIS & 400 \\
 & 8748 & 90 & P60/SEDM & 2250 & SN 2020aekp & 9224 & 19 & P60/SEDM & 2160 \\
 & 8761 & 103 & P200/DBSP & 600 & & 9342 & 132 & P60/SEDM & 2160 \\
SN 2019rvb & 8766 & 14 & P60/SEDM & 2250 & & 9343 & 132 & NOT/ALFOSC & 1200 \\
 & 8780 & 26 & P200/DBSP & 600  & & 9362 & 151 & P60/SEDM & 2700 \\
SN 2020onv & 9058 & 23 & P60/SEDM & 1800  & & 9381 & 169 & NOT/ALFOSC & 2400 \\
 & 9062 & 27 & P60/SEDM & 1800  & & 9404 & 191 & P60/SEDM & 2700 \\
 & 9069 & 33 & P60/SEDM & 1800  & & 9425 & 211 & NOT/ALFOSC & 1800 \\
 & 9070 & 34 & LT/SPRAT & 750  & & 9434 & 220 & P60/SEDM & 2700 \\
 & 9073 & 37 & P60/SEDM & 1800  & & 9448 & 233 & P60/SEDM & 2700 \\
 & 9074 & 38 & NOT/ALFOSC & 450  & & 9468 & 252 & P60/SEDM & 2700 \\
SN 2020qxz & 9076 & 13 & P60/SEDM & 2250  & & 9569 & 348 & P60/SEDM & 2700 \\
 & 9087 & 22 & P60/SEDM & 2250  & & & & & \\
 & 9092 & 26 & NOT/ALFOSC & 1800  & & & & & \\
 & 9098 & 32 & P60/SEDM & 2250  & & & & & \\
 & 9101 & 34 & NOT/ALFOSC & 1200  & & & & & \\
 & 9107 & 40 & P200/DBSP & 900  & & & & & \\
 & 9112 & 45 & Keck1/LRIS & 300  & & & & & \\
 & 9121 & 53 & P60/SEDM & 2250  & & & & & \\
 & 9141 & 71 & Keck1/LRIS & 399  & & & & & \\
 &      &    &            &      & & & & & \\
 &      &    &            &      & & & & & \\
 \bottomrule
    \end{tabularx}
    \label{tab:samplespec}
\end{table*}

\section{Analysis} 

\subsection{Photometry}
\subsubsection{Explosion epoch estimates}
For the purpose of this paper, the `explosion time' simply refers to the time when optical flux rises above the zero-point baseline (i.e. first light). We used pre-peak $g,r,i$-band ZTF photometry and $c,o$-band ATLAS photometry (binned in 1-day bins), when available, for our analysis. For each SN, the light curve was interpolated using Gaussian process regression to obtain the peak flux epoch, then a power-law (PL) model was fit using epochs from baseline to 60\% of peak brightness in each band following \citet{Miller2020}. The PL fits converged in at least one band for 6 out of 12 BTS SNe Ia-CSM. For the rest, we simply took the middle point between the first $5\sigma$ detection and the last upper limit before this detection as the explosion epoch with half of the separation between these two points as the uncertainty. 

The explosion time estimates, light curve bands used for the PL fits and the $1\sigma$ uncertainties on explosion times are listed in Table~\ref{tab:exptime}. The unfilled `PL fit filters' column in the table are the SNe for which the PL fit did not converge and averages were used. For the PL fits this typically constrains the time of explosion to within a fraction of a day. Given the high cadence of the ZTF survey, even in the cases where we use only the last non-detection the uncertainty range is typically less than 3 days. Only for SN 2020uem is the date of explosion virtually unconstrained ($\pm57$ days) as it was behind the sun at the time of explosion. 

Although for SN 2019ibk the explosion time is formally constrained with a $\pm3$ day uncertainty, this estimate was derived using only ATLAS $o$-band data right after the SN emerges from behind the sun. There is not a clear rise observed over a few epochs but two non-detections before a 5$\sigma$ detection. It is possible that the actual peak of this SN occurred earlier while it was behind the sun and the rising $o$-band points after it emerged are due to a second peak or bump (similar to SN 2018evt, in that case the actual rise was caught before the SN went behind the sun in ASAS-SN data). If the former explosion epoch estimate from $o$-band is to be believed then SN 2019ibk would be the most sub-luminous among the SNe Ia-CSM, peaking at $-17.5$.

\begin{table}[]
    \centering
    \caption{Explosion time epoch estimates derived from pre-peak multi-band light curves. For 6 out of 12 SNe Ia-CSM, we were able to fit a power-law model to multi-band data following \citet{Miller2020}. For the remaining 6 SNe, the explosion epoch was estimated by taking the mean of the first $5\sigma$ detection and last upper-limit before the first detection.}
    \begin{tabular}{c c c c}
    \hline
    \hline
    IAU Name & PL fit filters & $t_o$ & $1\sigma$ interval \\
     & & (MJD) & (days) \\
    \hline
    SN 2018crl & $g,r,o$ & 58271.83 & [$-$0.48,$+$0.38] \\ 
    SN 2018gkx & $r,o$ & 58371.34 & [$-$0.64,$+$0.53] \\ 
    SN 2018evt & - & 58334.26 & [$-$2.00,$+$2.00] \\ 
    SN 2019agi & - & 58502.48 & [$-$1.51,$+$1.51] \\ 
    SN 2019ibk & - & 58654.61 & [$-$2.99,$+$2.99] \\ 
    SN 2019rvb & $g,r,i,o$ & 58749.16 & [$-$0.79,$+$0.60] \\ 
    SN 2020onv & $o$ & 59032.75 & [$-$2.49,$+$1.10] \\ 
    SN 2020qxz & $g,r,o$ & 59063.05 & [$-$0.51,$+$0.45] \\ 
    SN 2020uem & - & 59117.03 & [$-$56.63,$+$56.63] \\ 
    SN 2020xtg & - & 59130.14 & [$-$0.04,$+$0.04] \\ 
    SN 2020abfe & $g,r,o$ & 59159.36 & [$-$2.16,$+$2.23] \\ 
    SN 2020aekp & - & 59204.53 & [$-$5.50,$+$5.50] \\ 
    \hline
    \end{tabular}
    \label{tab:exptime}
\end{table}
\subsubsection{Duration and absolute magnitudes}

\begin{figure}[!htbp]
    \centering
    \includegraphics[width=0.48\textwidth]{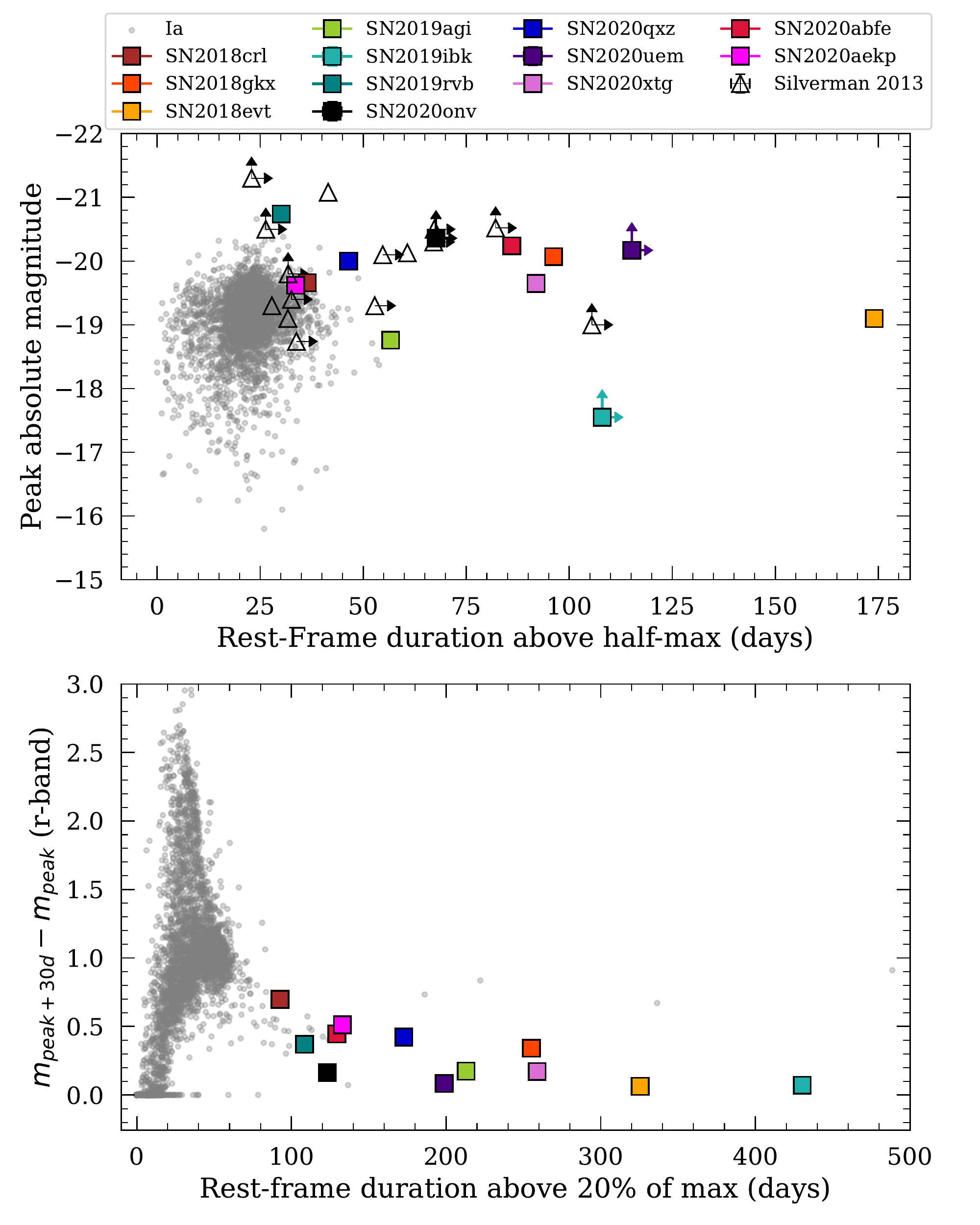}
    \caption{\textit{Top:} Location of our 12 SNe Ia-CSM in the peak absolute magnitude vs. rest-frame duration above half max phase space. The colored points are the BTS SNe Ia-CSM and the gray points are the rest of the BTS SNe Ia. Also shown with empty triangles are the SNe Ia-CSM from S13. The vertical arrows mark the upper limits to peak absolute magnitudes and horizontal arrows mark the lower limits to durations of SNe not having pre-peak coverage. \textit{Bottom:} Change in magnitude 30 days after peak ($\Delta m_{30}$) vs. rest-frame duration above 20\% of peak-flux for BTS SNe Ia and SNe Ia-CSM. These criteria were used to filter out potential SNe Ia-CSM from all SNe Ia and demonstrate that SNe Ia-CSM occupy a distinct portion in this phase space. However some gray points (not SN Ia-CSM) remain on the longer duration side and are the false positive cases described in \S2.1.}
    \label{fig:phasespace}
\end{figure}

Figure~\ref{fig:phasespace} shows the SNe Ia-CSM (colored squares) in our sample in the duration-luminosity and duration-$\Delta m_{30}$ phase space. In the top panel, the x-axis is duration above half-max and the y-axis is the peak absolute magnitude (see Table~\ref{tab:prop}) when we have photometric coverage both pre-peak and post-peak. For SNe missing the pre-peak coverage, their discovery magnitude is taken to be the upper limit to peak absolute magnitude and the duration from discovery the lower limit to duration above half-max (marked by arrows in Figure~\ref{fig:phasespace}). The BTS SN Ia sample is shown in gray points, and we also show the SNe Ia-CSM presented in S13 with empty triangles for comparison in the top panel. In the bottom panel, the x-axis is duration above 20\% of peak flux ($\Delta t_{20}$) and the y-axis is $\Delta m_{30}$, the two parameters used in the selection criteria. Most of the SNe Ia-CSM lie on the longer duration and brighter luminosity side, and are even more distinctly separated in the $\Delta t_{20}$-$\Delta m_{30}$ phase space. This makes the SN initial decline rate and duration useful tools for identifying thermonuclear SNe potentially interacting with CSM, if they have not revealed themselves already in their early time spectra. The gray points lying in the same phase space as SNe Ia-CSM are the false positive cases described in \S2.1. Also worth noting is that the duration calculated by taking the flux above half of peak flux value does not capture the true duration of the light curve when the plateau phase falls below half-max as is the case for SN 2020aekp ($>500$ days light curve) but $\Delta t_{20}$ and $\Delta m_{30}$ do.

\subsubsection{Light and color curves}

We have good pre-peak coverage in ZTF data for 8 of the 12 SNe in our sample\footnote{except for SNe 2018evt, 2019ibk, 2020onv and 2020uem.}. SN 2018evt was discovered by ASAS-SN on JD 2458341.91 \citep{discovery2018TNStuhrs} and classified by ePESSTO the next day \citep{TNS2018evt}, around 115 days before the first detection in ZTF when the SN came back from behind the sun. Hence we have only one epoch of pre-peak photometry and one early spectrum for SN 2018evt.

Our mixed bag of SNe Ia-CSM show post-maximum decline rates ranging from 0.5 to 2.0\,mag\,100d$^{-1}$ in the $r$ band from peak to $\sim100$ days post peak. The median decline rate is 1.07\,mag\,100d$^{-1}$, which is much slower than the decline rates of normal SNe Ia. We see a variety of changes in decline rates after around 100 days from peak. Two SNe (2020onv and 2020abfe) show no change and have a constant slow decline throughout. Four SNe (2018gkx, 2019agi, 2019ibk and 2019rvb) evolve to a shallower slope going from $\sim0.6$--1\,mag\,100d$^{-1}$ to $\sim0.2$--0.5\,mag\,100d$^{-1}$. Three SNe (2018crl, 2020qxz and 2020aekp) show a major change in decline rate with the light curves becoming almost flat, and SN 2020aekp shifts back to a slow decline from this plateau after $\sim200$ days. In three cases, the decline rate actually becomes steeper, SN 2018evt goes from 0.52\,mag\,100d$^{-1}$ to 1.4\,mag\,100d$^{-1}$, SN 2020uem goes from 0.52\,mag\,100d$^{-1}$ to 1.25\,mag\,100d$^{-1}$ and SN 2020xtg seems to go from 0.61\,mag\,100d$^{-1}$ to 1.35\,mag\,100d$^{-1}$ (even though there is only one epoch at late times to measure this change). The 3 SNe with fastest initial decline rates ($\gtrsim1.5$\,mag\,100d$^{-1}$ in the $r$ band) are similar to SN 2002ic (initial decline of 1.66\,mag\,100d$^{-1}$ in $V$) and PTF11kx (initial decline of 3.3\,mag\,100d$^{-1}$ in $R$) and coincidentally are also the ones that evolve into a plateau. The rest of the sample have initial decline rates comparable to SN 1997cy (0.75\,mag\,100d$^{-1}$) and SN 2005gj (0.88\,mag\,100d$^{-1}$) \citep{inserra2016}. From these observations, we can conclude that SNe Ia-CSM exhibit a range of slow evolution indicating that there exists a continuum of phases at which strong CSM interaction begins to dominate the powering of the light curves for these SNe. It is, however, difficult to pinpoint the exact phase when interaction starts from the light curve without modeling. CSM interaction could be affecting the peak brightness significantly even in cases where interaction only appears to dominate after a few weeks (SNe 2018crl, 2020qxz 2020aekp). Considering the average peak phase to be $\sim20$ days past explosion from the light curves and assuming an ejecta velocity of $\sim20000$\,km\,s$^{-1}$, the CSM is located at $\sim3.5\times10^{15}$\,cm. This estimate can be refined by considering the phase of the earliest spectrum that shows interaction signatures (see \S\ref{sec:specanalysis}). At late times, all the decline rates are slower than that expected from Cobalt decay (0.98\,mag\,100d$^{-1}$), confirming that the power from CSM interaction dominates the light curve behaviour for a long time.

\begin{figure*}[h]
    \centering
    \includegraphics[width=0.98\textwidth]{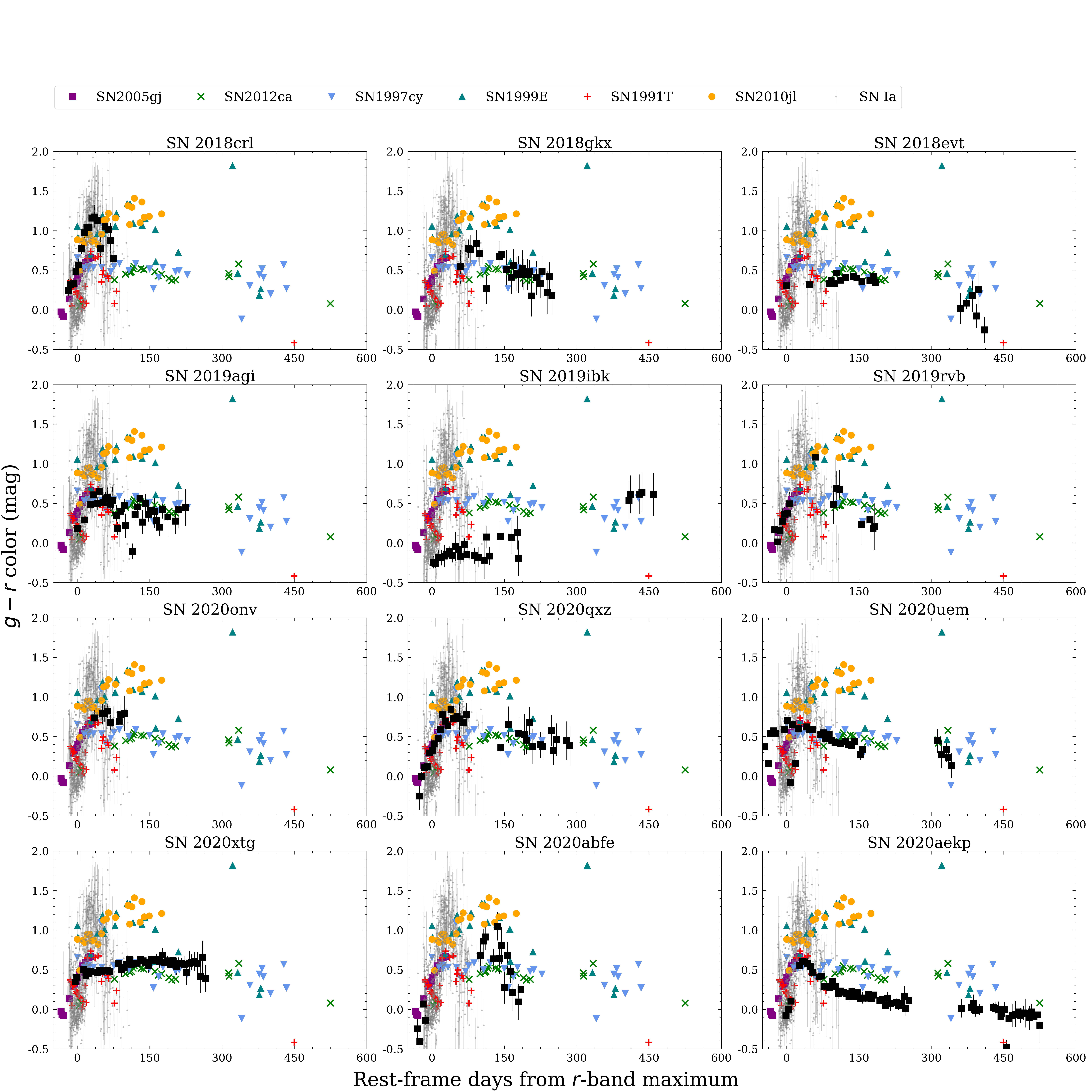}
    \caption{Color evolution ($g-r$) of BTS SNe Ia-CSM from $r$-band maximum (plotted in black) compared with SNe 2005gj, 1997cy, 1999E (Ia-CSM), SN 2012ca (IIn/Ia-CSM), SN 2010jl (IIn), SN 1991T (SN Ia) and ZTF SNe Ia (gray lines). As can be seen for up to $\sim150$ days, our SNe Ia-CSM tend to be redder than SNe Ia and at late times develop a plateau similar to other interacting SNe (IIn/Ia-CSM).}
    \label{fig:color}
\end{figure*}

Figure~\ref{fig:color} shows the $g-r$ color evolution 
of our sample SNe as a function of phase (rest-frame days from $r$-band maximum), comparing them with some famous SNe Ia-CSM (SNe 2005gj, 1997cy, 1999E), and SNe 2012ca (Ia-CSM/IIn), 2010jl (IIn) and 1991T (over-luminous Type Ia). The color evolution of normal SNe Ia from ZTF \citep{Dhawan2022} is shown in grey lines. We use $g-r$ colors when available, otherwise we estimate the $g-r$ color by fitting Planck functions to estimate the black-body temperatures from the $V-R$ colors. Our SNe Ia-CSM show similar color evolution as the older Type Ia-CSM/IIn interacting SNe, i.e. the $g-r$ color increases gradually for about 100 days and then settles onto a plateau or slowly declines, and one object (SN 2019ibk) becomes redder at late times similar to SN 2012ca. The interacting SNe are redder at late times compared to the normal SNe Ia. 

\subsubsection{Mid-IR brightness comparison}\label{midir}
\begin{figure}[!htbp]
    \centering
    \includegraphics[width=0.48\textwidth]{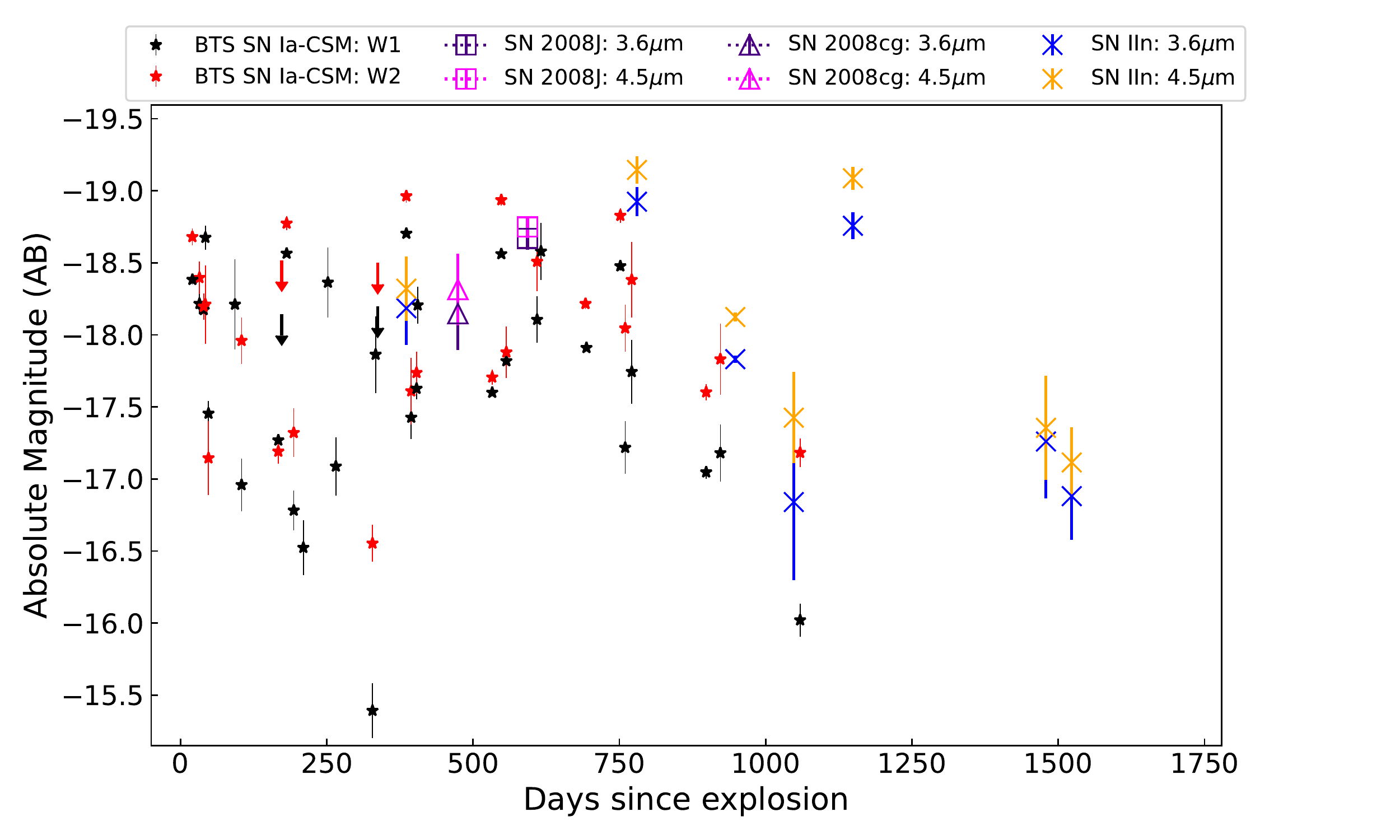}
    \caption{unWISE detections in the W1 and W2 bands of BTS SNe Ia-CSM. The W1 and W2 points are marked with black and red filled stars respectively. Spitzer IRAC photometry of SNe IIn (blue and orange crosses) and two SNe Ia-CSM from \citet{fox2011} (SNe 2008cg and 2008J in empty triangle and square) are also shown for comparison. 9 out of 12 BTS SNe Ia-CSM are as bright in mid-IR as other interacting SNe ($\sim-16$ to $\sim-19$). The upper limits for SN 2020abfe are shown in black and red filled upside down triangles.}
    \label{fig:wiselc}
\end{figure}

Out of 12 SNe in our sample, only one observed (SN 2020abfe) did not have 3$\sigma$ detections post explosion in the unWISE difference photometry light curves and two (SNe 2019rvb and 2020qxz) did not have coverage post explosion. The unWISE light curves for the rest of the SNe Ia-CSM having $>3\sigma$ detections in W1 (3.3~$\mu$m) and W2 (4.6~$\mu$m) bands are shown in Figure~\ref{fig:wiselc} (black and red stars) along with \textit{Spitzer} IRAC survey data of SN 2008cg (indigo and magenta empty triangles), SN 2008J (indigo and magenta empty squares) (both Ia-CSM) and some SNe IIn (blue and orange crosses) taken from \citet{fox2011}. The most nearby SN in our sample, SN 2018evt, is among the brightest ($\sim17$ AB mag) in MIR at least until $\sim$1000 days after explosion and has a bumpy light curve. SNe 2019ibk and 2018crl however are the most luminous with an absolute magnitude of $-18.7$ mag in the W1 band. The brightness of the BTS SNe Ia-CSM is comparable with other interacting SNe and span a similar range ($-16$ to $-19$). However, SNe IIn have been detected until even later epochs (up to 1600 days) than SNe Ia-CSM, probably due to the larger number of SNe IIn at closer distances. SN 2020abfe has upper limits around $\sim-18$ in W1 band and $\sim-18.5$ in W2 band up to $\sim$300 days post explosion shown with upside down filled triangles. As the mid-IR luminosity can be fainter than these limits for SNe Ia-CSM (as can be seen for other nearby SNe in this sample) and SN 2020abfe is at a redshift of 0.093, it might just be out of reach for WISE.

This brightness of SNe Ia-CSM in mid-IR can be indicative of existing or newly formed dust. A clear signature of new dust is reduced flux in the red wing of the H$\alpha$ emission line at late phases as the new dust formed in the cold dense shell behind forward shock absorbs the far-side (redshifted) intermediate and narrow line emission (see bottom panel of Fig.~\ref{fig:speccomp}). For our sample, this reduction in H$\alpha$ red wing is the most pronounced for SN 2018evt. 

\subsubsection{Bolometric luminosity}
As the SN Ia-CSM luminosity is dominated by CSM interaction, their spectra comprise of a pseudo-continuum on the blue side and strong H$\alpha$ emission on the red side, hence a blackbody fit to multi-band photometric data is not appropriate to estimate the bolometric luminosity. Instead we calculate a pseudo-bolometric luminosity from the available multi-band optical data by linearly interpolating the flux between the bands and integrating over the optical wavelength range spanned by the ATLAS and ZTF bands. The individual band light curves are first interpolated using Gaussian process regression to fill in the missing epochs. This estimate places a strict lower limit on the bolometric luminosity.

\begin{figure}[!htbp]
    \centering
    \includegraphics[width=0.48\textwidth]{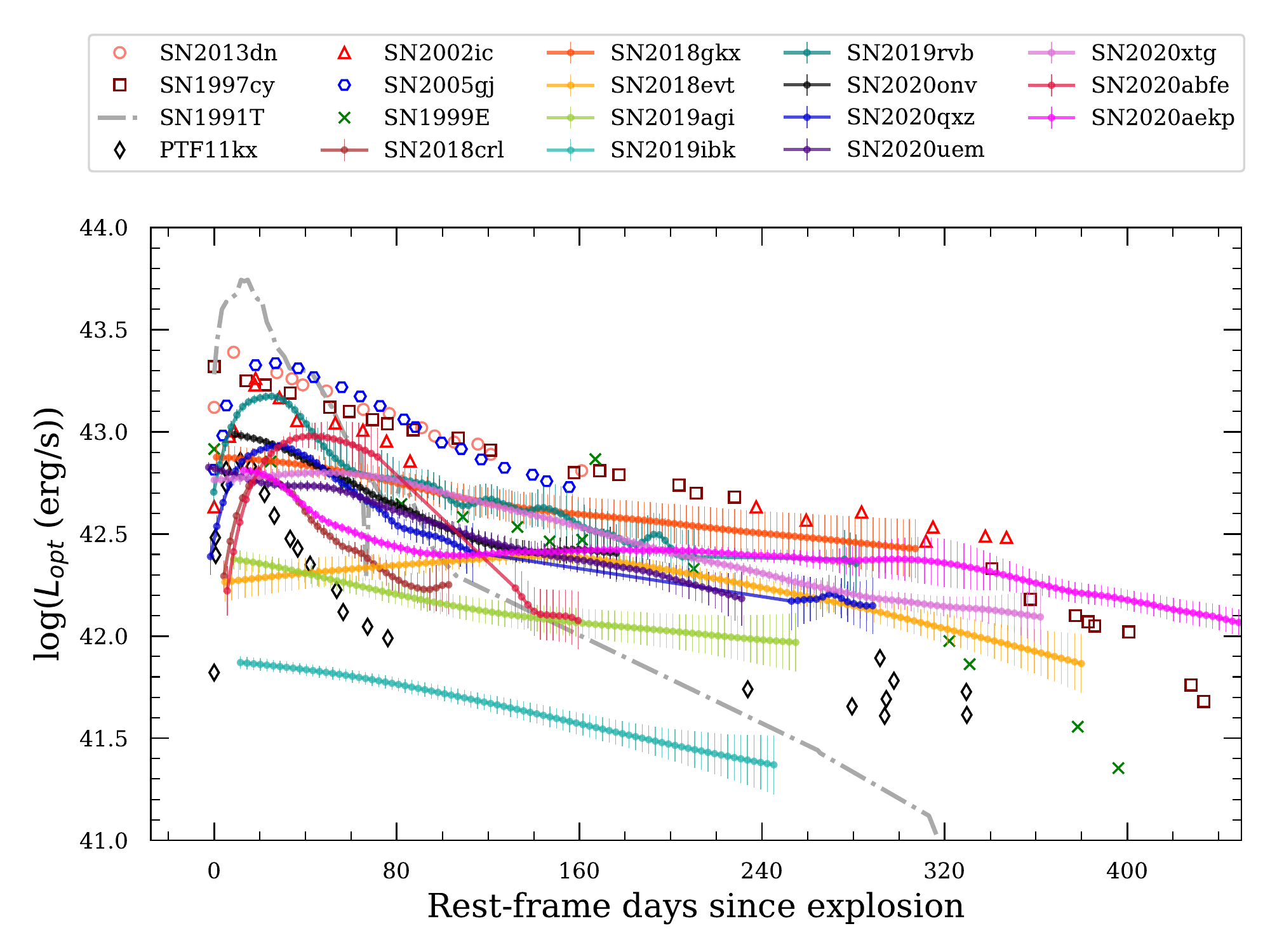}
    \caption{Pseudo-bolometric luminosity light curves of BTS SNe Ia-CSM compared with pseudo-bolometric light curves of SNe 1991T, 1997cy, 1999E, 2002ic, 2005gj, 2013dn, and PTF11kx from literature. The light curves in each filter having more than 10 epochs were interpolated using Gaussian process regression to fill in the missing epochs, and at each epoch the fluxes between the bands were linearly interpolated and integrated over the optical wavelength range spanned by ZTF and ATLAS filters to get the pseudo-bolometric luminosity. For BTS SNe, the phases are with respect to the estimated explosion epochs, while for comparison SNe the phases are with respect to discovery.
    }
    \label{fig:bololum}
\end{figure}

In Figure~\ref{fig:bololum} we show the pseudo-bolometric luminosity of our SN Ia-CSM sample in comparison with SN 1991T (Type Ia), SNe 1997cy, 1999E, 2002ic, 2005gj, 2013dn and PTF11kx (Ia-CSM). Multi-band photometric data were taken from the  Open Supernova Catalog \citep{opensncatalog2017} for SN 1991T \citep{Fili1992,Ford1993,Schmidt1994} to generate the bolometric luminosity light curve through black body fitting. The pseudo-bolometric luminosity light curve for SN 1997cy was obtained from \citet{Germany2000}, for SN 2013dn from \citet{fox2015} and for SNe 2002ic, 2005gj, 1999E and PTF11kx from \citet{inserra2016}. 

All BTS SNe Ia-CSM show a slow evolution in bolometric luminosity, inconsistent with the decay of $^{56}$Co to $^{56}$Fe. The sample's overall luminosity decline rates are comparable to those of SNe 1997cy and 2013dn, as shown in Figure~\ref{fig:bololum}. Only SNe 2018crl and 2020aekp seem to show early decline in their pseudo-bolometric light curves similar to SN 1991T for about 40 days after peak like SN 2002ic and PTF11kx. Another BTS interacting SN Ia, ZTF20aatxryt \citep{Kool}, was found to follow the PTF11kx light-curve evolution very closely and as its light curve fell into a plateau the SN started showing signs of interaction with a helium-rich CSM and evolved into a helium-rich SN Ia-CSM. We have excluded ZTF20aatxryt from the sample as we focus on typical SNe Ia-CSM interacting with hydrogen-rich CSM in this study. At late phases ($\sim300$ days), the SNe Ia-CSM are approximately 100 times brighter than normal SNe Ia at the same epoch. Therefore, at these late phases, the luminosity and spectral features of SNe Ia-CSM are entirely dominated by CSM-interaction with little emergent SN flux. From the pseudo-bolometric light curves, we place a lower limit on the total radiated energy for SNe Ia-CSM to be 0.1--1.5 $\times 10^{50}$erg. 
This is well below the thermonuclear budget (E$_{kin}\sim10^{51}$ erg), but as this is a lower limit and some SNe in the sample have unconstrained peaks, the true total radiative energy might come close to the thermonuclear budget, requiring high conversion efficiency to achieve their luminosity.

\subsection{Spectroscopy} \label{sec:specanalysis}
\begin{figure*}[!htbp]
    \centering
    \includegraphics[width=0.98\textwidth]{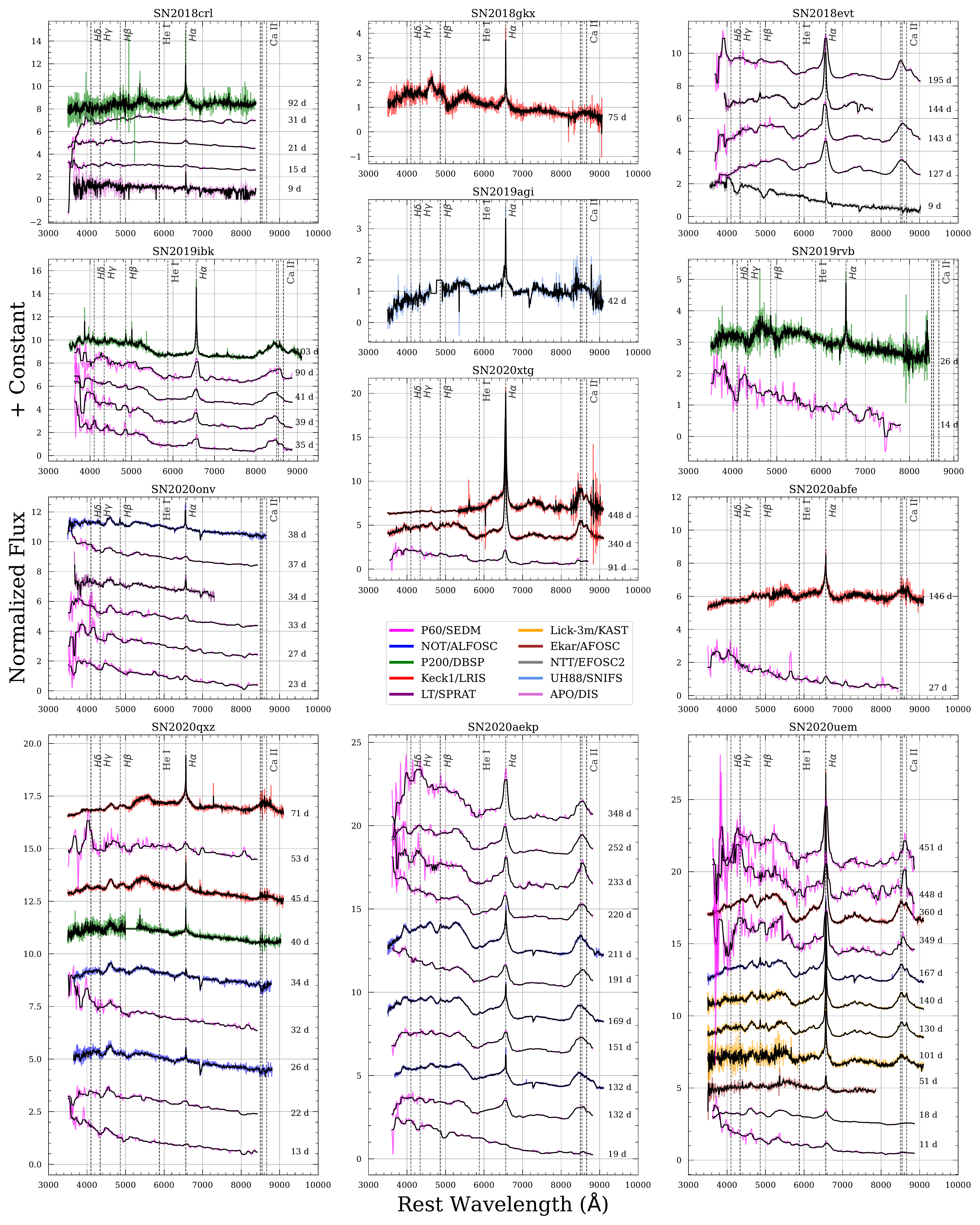}
    \caption{Spectral series of all SNe Ia-CSM presented in this paper. The rest-frame phases are shown alongside the spectra in each subplot and have been calculated using the explosion epoch estimate. The colors depict different instruments used to obtain this data. Major emission lines are marked with vertical dashed lines.}
    \label{fig:specseries}
\end{figure*}

Figure~\ref{fig:specseries} displays the spectral series obtained for the BTS SNe Ia-CSM. Most of the early time spectra were taken with the SEDM, the BTS workhorse instrument (R $\sim$100), which is not able to resolve the narrow CSM lines. Therefore, these SNe were followed up with higher resolution instruments to get more secure classifications. For each spectrum in Figure~\ref{fig:specseries}, the phase is provided with respect to the explosion epoch estimate given in Table~\ref{tab:exptime}. We have spectra ranging from a few to around 470 days from explosion. Considering the well constrained explosion times of SN 2018evt, presence of narrow H$\alpha$ in its first spectrum at 8 days since  explosion and assuming a typical ejecta velocity of $\sim$20000\,km\,s$^{-1}$, this implies that the CSM interaction start as close as $\sim$1.4$\times10^{15}$\,cm.

\begin{figure*}
    \centering
    \includegraphics[width=0.95\textwidth]{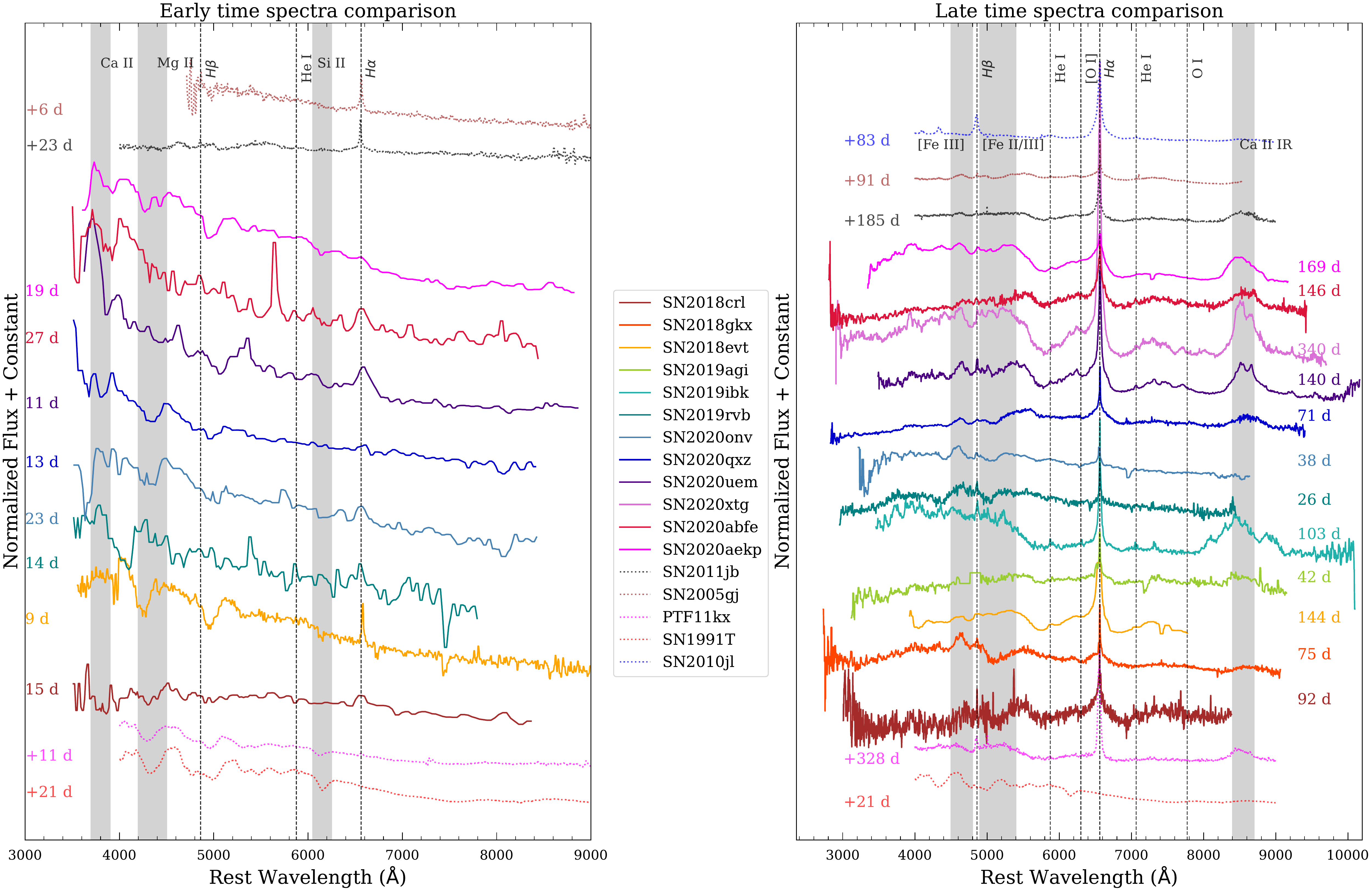}
    \includegraphics[width=0.95\textwidth]{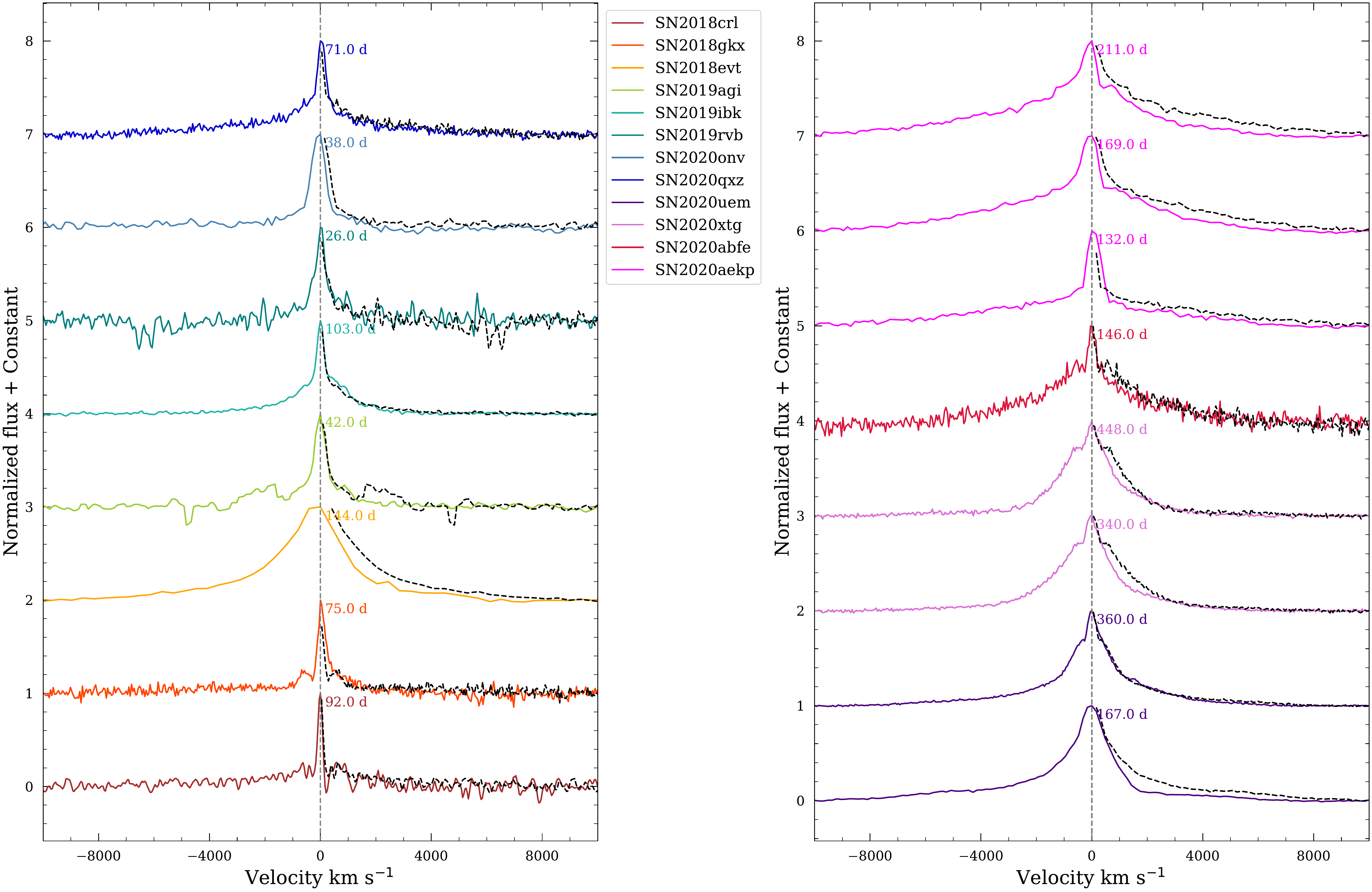}
    \caption{Top left: Early-time spectra of BTS SNe Ia-CSM with phases between 0 and 30 days since explosion compared to spectra of SNe 2011jb, 2005gj, 1991T and PTF11kx (phases in days since discovery). Top right: Late-time spectra of BTS SNe Ia-CSM (phases ranging from 40 to 370 days since explosion) compared to spectra of SNe 2011jb, 2005gj, 2010jl and PTF11kx (phases in days since discovery).} Bottom left and right: H$\alpha$ line profiles (post continuum removal) with the blue side reflected across the peak flux, marked by dashed lines. SNe 2020aekp, 2020abfe, 2020xtg and 2020uem in the right panel, and SNe 2018crl, 2018gkx, 2018evt, 2019agi, 2019ink, 2019rvb, 2020onv, 2020qxz in left panel.
    \label{fig:speccomp}
\end{figure*}

Figure~\ref{fig:speccomp} shows the early time (left) and late time (right) spectral behaviour of the BTS SNe Ia-CSM together with a few historical SNe for comparison, namely SNe Ia-CSM SN 2011jb \citep{Silverman_2013}, SN 2005gj and PTF11kx, the Type Ia SN 1991T and the well-observed  Type IIn SN 2010jl. Vertical gray regions mark typical SN Ia absorption features and [Fe II/III] line regions, and vertical dashed lines mark the Balmer emission lines. The sample spectra have been multiplied by a constant factor to magnify relevant spectral features. In the following paragraphs, we compare the observations of some of the spectral features with previous analysis of this class  \citep[][]{Silverman_2013,fox2015,inserra2016}.

A few of our early time SNe Ia-CSM show underlying SN Ia absorption features like PTF11kx and SN 2002ic (most are, however, quite diluted and also affected by the low resolution and signal-to-noise ratio (SNR) of the SEDM spectra), the most notable being SNe 2018evt, 2020qxz and 2020aekp. SNe 2020qxz and 2020aekp also have among the fastest initial post-peak decline rates in the sample, similar to PTF11kx, while coverage around peak is not available for SN 2018evt. On the other hand, SNe with slower decline rates similar to SN 1997cy and SN 2005gj have more SN IIn-like early time spectra dominated by blue pseudo-continuum and Balmer emission. The faster decline rate suggests we are still seeing some of the emission from the ejecta at those phases. To unveil the nature of the progenitor of interacting SNe, it is therefore necessary to obtain some spectroscopic follow-up before peak light. Spectroscopic data at the phase of transition to interaction-dominated luminosity would also help in deducing the extent and density structure of the optically thick CSM.

Late time spectra of SNe Ia-CSM look very similar to those of SNe IIn, heavily dominated by H$\alpha$ emission. The CSM interaction masks the underlying SN signature and we instead see late-time spectra riddled with photoionized CSM lines. In some cases, the photosphere might lie in an optically thick cold dense shell (CDS) formed between the forward and reverse shocks which obscures the ejecta completely \citep{smith2008,chugai2004}. The continuum is also enshrouded under a blue quasi-continuum from a forest of iron-group element lines (S13) as identified and analyzed for SNe 2012ca and 2013dn by \citet{fox2015}. 

The blue quasi-continuum blend of iron lines ([\ion{Fe}{3}] lines around $\sim$4700\,\AA\ and [\ion{Fe}{2}] around $\sim$5200\,\AA) in the spectra of the BTS SN Ia-CSM sample (see Figure~\ref{fig:speccomp} top right panel) is the dominant feature blue-ward of 5500\,\AA\, but the ratio of [\ion{Fe}{3}]/[\ion{Fe}{2}] is much weaker compared to for SNe Ia (like SN 1991T). This feature is more apparent in the SNe Ia-CSM like PTF11kx and SN 2002ic that became interaction-dominated later than for other SNe Ia-CSM such as SNe 1997cy, 1999E and SN 2012ca (SN Ia-CSM/IIn, for which a clear type has not been established). \citet{inserra2014} argues for a core-collapse origin for SN 2012ca given this low amount of [\ion{Fe}{3}] along with the detection of blueshifted Carbon and Oxygen lines (which however, were later argued to be [\ion{Fe}{2}] lines by \citealt{fox2015}). S13 instead argues in favor of a thermonuclear origin given the presence of this blue quasi-continuum, despite [\ion{Fe}{3}] being weaker. \citet{fox2015} points out that a similarly suppressed ratio of [\ion{Fe}{3}]/[\ion{Fe}{2}] is observed in some SNe Ia, particularly the super-Chandra candidate SN 2009dc, for which the explanation was suggested to be a low ionization nebular phase owing to high central ejecta density and low expansion velocities \citep{taubenberger2013}. \citet{fox2015} argue that in the case of SNe Ia-CSM, a lower ionization state could arise owing to the deceleration of ejecta by the dense CSM explaining the Fe line ratio suppression. Since Ca has lower first and second ionization potentials than Fe, the detection of [\ion{Ca}{2}] $\lambda\lambda$7291, 7324 would be consistent with this low ionization, which \citet{fox2015} confirms for SNe 2012ca and 2013dn. Indeed, we find clear evidence of [\ion{Ca}{2}] emission for 8 out of 12 SNe in our sample and moderate to weak signal for the remaining 4. Although this does favor the argument for a thermonuclear origin, a similar blue quasi-continuum is also observed in other interacting SN types like SNe Ibn (SN 2006jc, \citealt{Foley_2007}) and SNe IIn (SNe 2005ip and 2009ip), making Fe an incomplete indicator of the progenitor nature (see detailed discussion in \citealt{fox2015}).

We do not find strong evidence of \ion{O}{1} $\lambda7774$ or [\ion{O}{1}] $\lambda\lambda$6300, 6364 emission in our sample, although they might be present at very weak levels in some SNe (e.g.~SN 2020uem). SN 2020uem has strong emission lines at 6248, 7155 and 7720 \AA\ which are consistent with being iron lines and were also observed in SNe 2012ca, 2013dn and 2008J. S13 note that the very broad emission around 7400\,\AA\ can be due to a blend of [\ion{Ca}{2}] $\lambda\lambda$7291, 7324 and [\ion{O}{2}] $\lambda\lambda$7319, 7330, however we note that this broad emission is likely to be from calcium as \ion{O}{2} is harder to excite than \ion{O}{1} which is either very weak or absent in our spectra. The broad Ca NIR triplet feature resulting from electron scattering is the next strongest feature after the Balmer emission and is present in all mid to late-time spectra of the SNe in our sample where the wavelength coverage is available. We observe it increasing in relative strength with phase, at least for a year, after which we no longer have spectral coverage.

The bottom panel of Figure~\ref{fig:speccomp} shows the line profile of H$\alpha$, with the blue side reflected over the red side at the maximum flux after continuum removal. We do see evidence of diminished flux in the red wing of H$\alpha$ at late phases in some SNe (most notable in SNe 2018evt and 2020uem), which can indicate formation of new dust in the post-shock CSM. S13 claim to observe this for all non-PTF SNe Ia-CSM in their sample starting at $\sim$75--100 days, while for the PTF SNe Ia-CSM they do not have spectra available post that phase range. For some BTS SNe Ia-CSM, we also do not have spectra available post 100 days which limits any analysis of this phenomenon for a large enough sample.

The spectra were reduced and processed as outlined in \S \ref{sec:dataspec} for the emission line analysis, the results of which are described in the next section. We used only good SNR SEDM spectra and intermediate resolution spectra for line identification and analysis.

\subsubsection{H$\alpha$, H$\beta$ and \ion{He}{1} emission lines}

\begin{figure}
    \centering
    \includegraphics[width=0.48\textwidth]{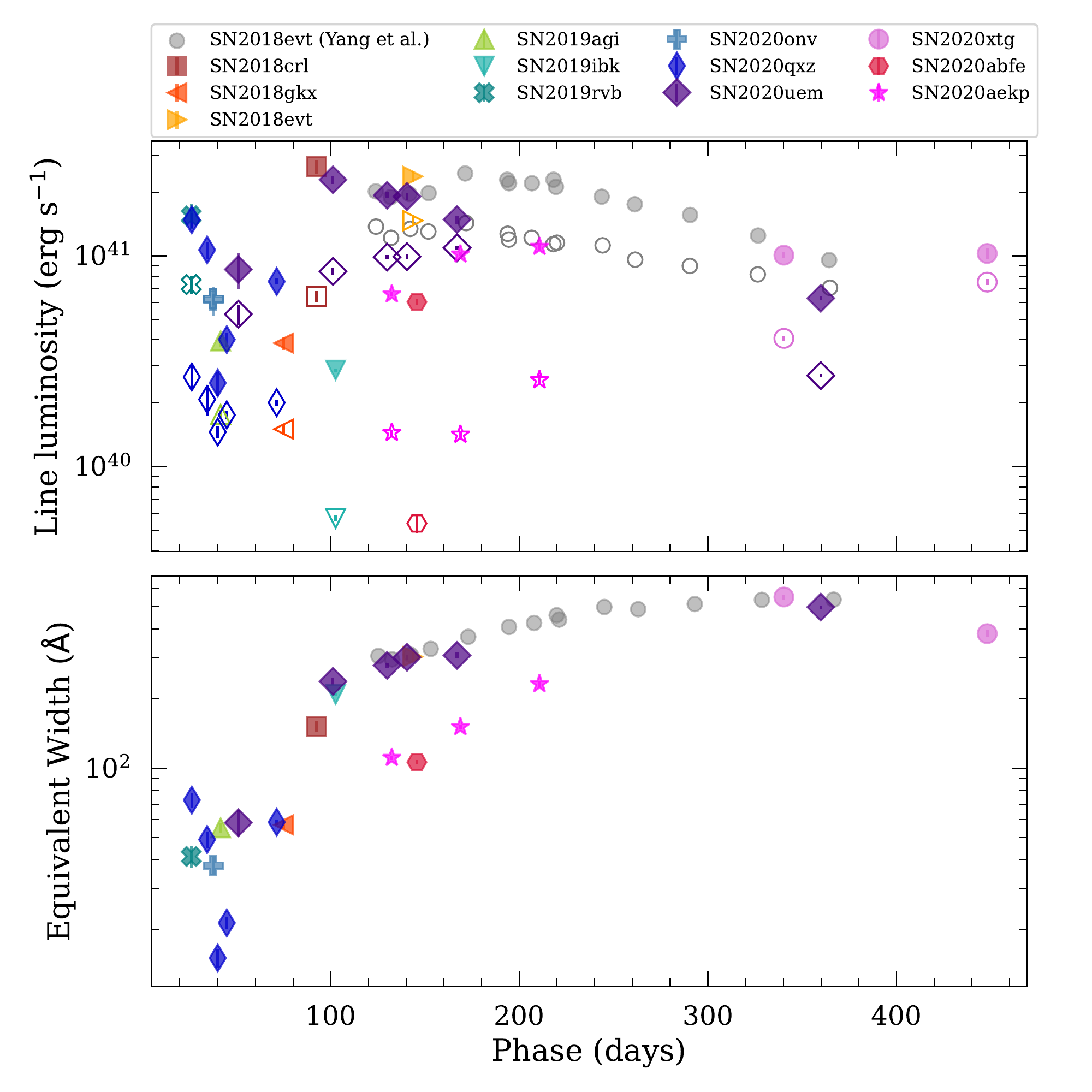}
    \caption{Integrated fluxes and equivalent widths of H$\alpha$ emission line with respect to SN phases for the BTS SN Ia-CSM sample. Broad component values are shown with filled markers and narrow component values with un-filled markers. SN 2018evt H$\alpha$ luminosities and EWs presented in \citet{Yang2022} are also shown in gray circles.}
    \label{fig:halpha}
\end{figure}

To analyze the H$\alpha$ line emission, we first fit the continuum level using the \texttt{fit\_continuum} function of the \texttt{specutils} Python package, where the continuum is estimated by a cubic function fitted on regions on each side of the line. We remove this continuum level and then fit the H$\alpha$ line with a broad and a narrow component Gaussian function using the \texttt{fit\_lines} function of \texttt{specutils} which returns the best fit Gaussian model and the $1\sigma$ uncertainty on the model parameters. We generate 1000 sample models within $1\sigma$ uncertainties of the parameters centered around the best-fit values and calculate the intensity, flux and velocity (FWHM) of the broad and narrow components for each model. Then we take the median and standard deviation of the intensity, flux and velocity FWHM distributions to get their final best value and $1\sigma$ uncertainty. The equivalent width was also calculated for the H$\alpha$ line using the model fit as well as directly from the data, and the difference between the values derived from model and data is reported as the error on the EW. All values are reported in Table~\ref{tab:halpha}. For 3 SNe in our sample, we have a series of intermediate resolution spectra through which we can trace the evolution of the H$\alpha$ line with phase. Figure~\ref{fig:halpha} shows this trend of the H$\alpha$ line parameters (integrated flux in the top panel and equivalent width in the bottom panel) versus phase for all SNe in our sample. The un-filled markers represent the narrow emission while the filled markers represent the broad emission. For SNe where this analysis could be done on multiple spectra, we see that the H$\alpha$ equivalent width generally increase over time, with some SNe showing fluctuations up to 100 days possibly due to interaction of ejecta with multiple CSM shells of varying density. For SN 2018evt, \citet{Yang2022} analyzed H$\alpha$ line properties from a comprehensive spectral series data, which are plotted in Figure~\ref{fig:halpha} in gray circles and seem to agree well with our analysis at comparable epochs.

\begin{figure}
    \centering
    \includegraphics[width=0.48\textwidth]{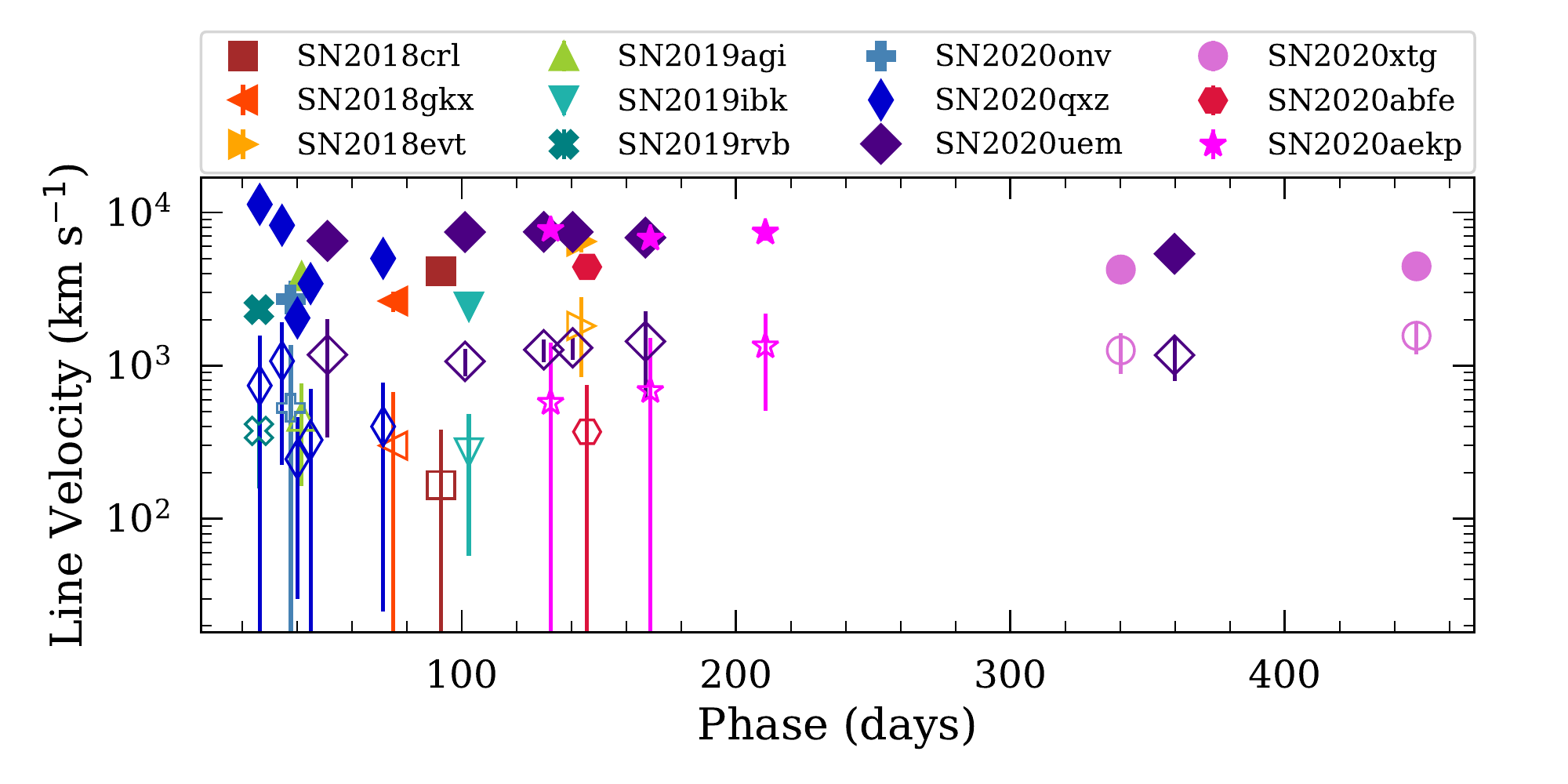}
    \caption{Velocity of H$\alpha$ emission line with respect to SN phases for the BTS SN Ia-CSM sample. Broad component values are shown with filled markers and narrow component values with un-filled markers.}
    \label{fig:halphavel}
\end{figure}

From the Gaussian profile line fitting analysis of the H$\alpha$ emission line, we found that the broader component has velocities ranging from $\sim$1000 to $\sim$4000\,km\,s$^{-1}$ (intermediate width) and the narrow component has velocities of about $\sim$200\,km\,s$^{-1}$ to $\sim$1000\,km\,s$^{-1}$ (see Figure~\ref{fig:halphavel}). The narrow component could only be resolved down to $\sim$300\,km\,s$^{-1}$ limited by the mediocre resolution of the spectrographs used (KeckI/LRIS R$\sim$800, P200/DBSP R$\sim$1000, NOT/ALFOSC has R$\sim$360). While we know that the narrow lines originate in the unshocked ionized CSM, the exact origin of the intermediate components is uncertain. They could arise from the post-shock gas behind the forward shock or from the shocked dense clumps in the CSM \citep{ChugaiDanziger1994}.

\begin{figure}
    \centering
    \includegraphics[width=0.48\textwidth]{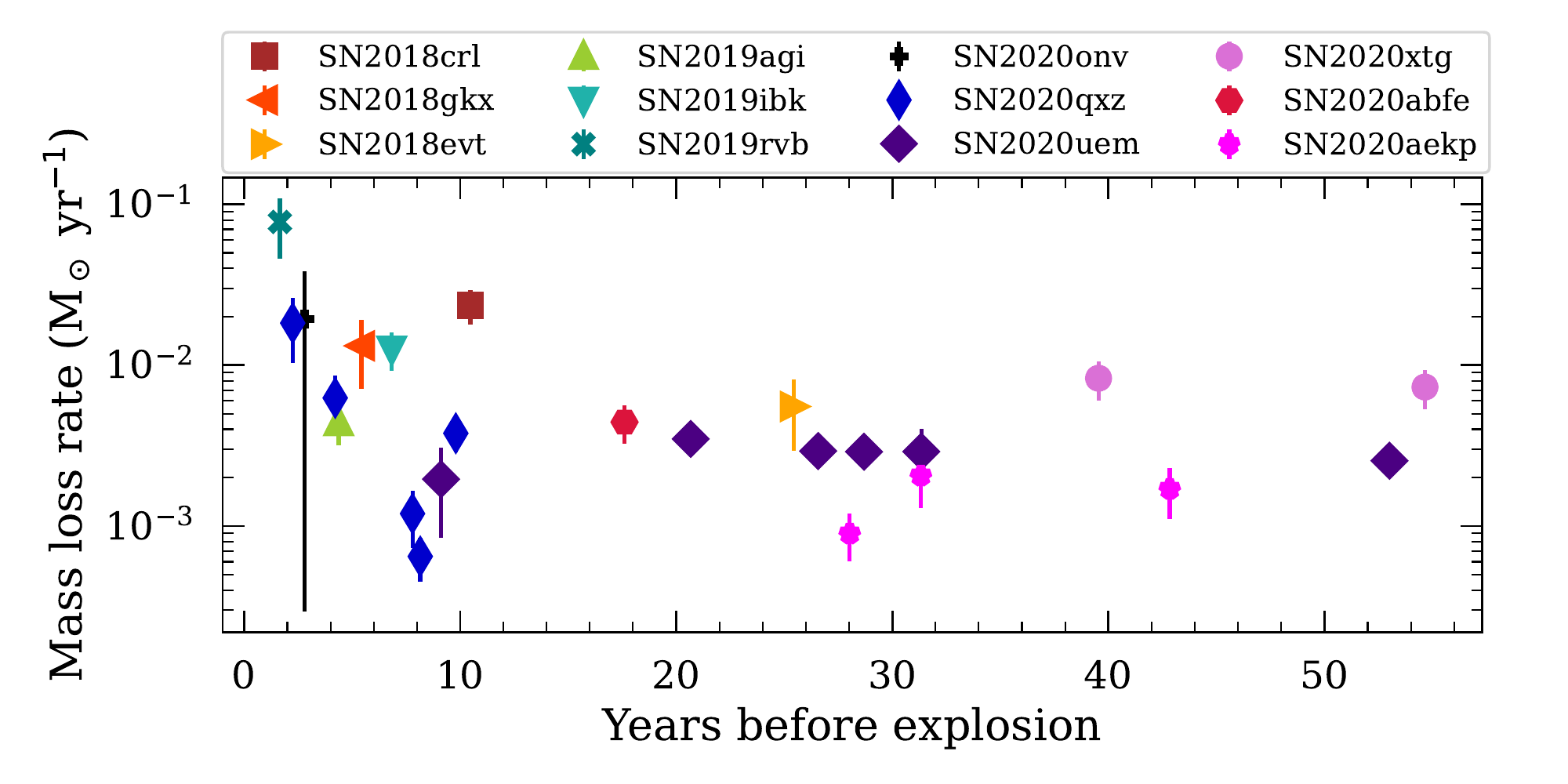}
    \caption{Mass-loss rates estimated from the luminosity of the broad component of H$\alpha$ for the BTS SNe Ia-CSM. A value of 100\,km\,s$^{-1}$ was assumed for the wind velocity.}
    \label{fig:massloss}
\end{figure}

The luminosities of the H$\alpha$ line measured from the BTS SNe Ia-CSM lie in the range 2.5--37$\times10^{40}$\,erg\,s$^{-1}$ which are comparable to the values from S13 who reported most of their SNe in the 1--10$\times10^{40}$\,erg\,s$^{-1}$ range except one object that had a luminosity of 39$\times10^{40}$\,erg\,s$^{-1}$. From the broad H$\alpha$ luminosity, we did a simple estimate of the mass-loss rate assuming spherically symmetric CSM deposited by a stationary wind $\rho \propto r^{-2}$ having velocity $v_w$ \citep{chugai1991,salamanca1998}. The mass-loss rate $\dot{M}$ can be related to the broad H$\alpha$ luminosity $L_{H\alpha}^{Broad}$ as \citep[their Eq. 2]{salamanca1998}
$$ L_{H\alpha}^{Broad} = \frac{1}{4}\epsilon_{H\alpha}\frac{\dot{M}}{v_w}v_s^3 $$
where $v_s$ is the shock velocity (obtained from the broad component velocity of the H$\alpha$ line). We used a value of 100\,km\,s$^{-1}$ considering previous high resolution spectral studies of SNe Ia-CSM \citep{Kotak2005,Aldering2006,dilday2012} for $v_w$ as we cannot fully resolve the narrow component and a maximum value of 0.1 for the efficiency factor $\epsilon_{H\alpha}$ \citep{salamanca1998}. The mass-loss rates were estimated from the available spectra and are shown in Figure~\ref{fig:massloss} as a function of years before explosion ($t_w = \frac{v_s t}{v_w}$, where $t$ is the phase of the spectra). For most SNe in the sample, the mass-loss rates lie between 0.001--0.02\,$M_\odot$\,yr$^{-1}$, except for SN 2019rvb which has $\sim$0.07\,$M_\odot$\,yr$^{-1}$ lost within 2 years prior the explosion. These rates are much higher than what could be attained from a red giant superwind ($\sim3\times10^{-4}$\,$M_\odot$\,yr$^{-1}$) but are comparable to previous estimates (calculated through multiple methods) for SNe Ia-CSM and require some unusual mechanism to reach such persistently higher mass-loss rates in the decades prior to explosion. Also to consider is that the simplistic assumption of spherical symmetry likely does not apply for SNe Ia-CSM. Evidence of multiple thin shells and asymmetric CSM was observed for PTF11kx \citep{dilday2012} and light curve modeling of SNe 1997cy and 2002ic suggested a better fit to a flat density profile rather than stationary wind \citep{ChugaiYungelson2004}. An asymmetric or clumpy CSM might be the norm for SNe Ia-CSM (and some SNe IIn) rather than the exception.

\begin{table*}[t]
    \centering
    \caption{Summary of H$\alpha$ line properties obtained from two-component Gaussian fitting.}
    \begin{tabular}{c|c|c|c|c|c|c}
    \hline
    \hline
\textbf{SN Name} & \textbf{Phase} & \textbf{Broad Flux} & \textbf{Narrow Flux} & \textbf{Total Flux} & \textbf{Broad Velocity} & \textbf{Narrow Velocity} \\
 & \small{(\text{days})} & \small{($10^{-16}$\,erg\,s$^{-1}$\,cm$^{-2}$)} & \small{($10^{-16}$\,erg\,s$^{-1}$\,cm$^{-2}$)} & \small{($10^{-16}$\,erg\,s$^{-1}$\,cm$^{-2}$)} & \small{FWHM (km\,s$^{-1}$)} & \small{FWHM (km\,s$^{-1}$)} \\
 \hline
SN 2018crl & 92 & 135.4$\pm$10.0 & 32.8$\pm$2.0 & 168.2$\pm$12.0 & 4137$\pm$312 & $<214$ \\
SN 2018gkx & 75 & 9.9$\pm$0.7 & 3.9$\pm$0.2 & 13.7$\pm$0.9 & 2640$\pm$398 & $<375$\\
SN 2018evt & 144 & 2020.3$\pm$128.5 & 1247.4$\pm$52.8 & 3267.7$\pm$181.3 & 6465$\pm$997 & 1816$\pm$973 \\ 
SN 2019agi & 42 & 52.7$\pm$3.6 & 23.7$\pm$1.1 & 76.4$\pm$4.7 & 3836$\pm$349 & 464$\pm$301 \\ 
SN 2019ibk & 103 & 85.6$\pm$1.7 & 17.0$\pm$0.5 & 102.6$\pm$2.3 & 2431$\pm$217 & 272$\pm$214 \\ 
SN 2019rvb & 26 & 22.0$\pm$3.0 & 10.4$\pm$1.0 & 32.5$\pm$4.1 & 2321$\pm$298 & 374$\pm$216 \\ 
SN 2020onv & 38 & 32.8$\pm$5.2 & 33.3$\pm$2.0 & 66.1$\pm$7.2 & 2714$\pm$879 & $<$834 \\ 
SN 2020qxz & 26 & 76.6$\pm$6.2 & 13.8$\pm$1.7 & 90.4$\pm$7.9 & 11294$\pm$1106 & $<836$\\
SN 2020qxz & 34 & 55.1$\pm$5.0 & 10.8$\pm$1.8 & 65.9$\pm$6.8 & 8252$\pm$1039 & 1070$\pm$845 \\ 
SN 2020qxz & 40 & 12.9$\pm$1.7 & 7.6$\pm$0.5 & 20.5$\pm$2.2 & 2049$\pm$284 & 245$\pm$215 \\ 
SN 2020qxz & 45 & 20.7$\pm$1.6 & 9.1$\pm$0.4 & 29.8$\pm$2.1 & 3429$\pm$419 & $<375$\\
SN 2020qxz & 71 & 39.1$\pm$1.3 & 10.4$\pm$0.4 & 49.5$\pm$1.7 & 5013$\pm$395 & 400$\pm$375 \\ 
SN 2020uem & 51 & 246.3$\pm$47.2 & 151.1$\pm$16.8 & 397.4$\pm$64.0 & 6520$\pm$1163 & 1178$\pm$840 \\ 
SN 2020uem & 101 & 655.2$\pm$28.9 & 241.2$\pm$9.6 & 896.4$\pm$38.4 & 7456$\pm$309 & 1066$\pm$217 \\ 
SN 2020uem & 130 & 552.9$\pm$17.6 & 281.8$\pm$6.2 & 834.8$\pm$23.8 & 7465$\pm$265 & 1269$\pm$215 \\
SN 2020uem & 140 & 545.4$\pm$20.0 & 283.4$\pm$6.8 & 828.8$\pm$26.7 & 7457$\pm$275 & 1308$\pm$216 \\ 
SN 2020uem & 167 & 424.3$\pm$19.0 & 312.0$\pm$7.7 & 736.3$\pm$26.6 & 6852$\pm$854 & 1439$\pm$834 \\ 
SN 2020uem & 360 & 179.8$\pm$4.0 & 77.4$\pm$1.4 & 257.2$\pm$5.4 & 5377$\pm$382 & 1170$\pm$375 \\ 
SN 2020xtg & 340 & 129.2$\pm$4.2 & 52.1$\pm$1.6 & 181.3$\pm$5.8 & 4242$\pm$382 & 1258$\pm$376 \\ 
SN 2020xtg & 448 & 131.7$\pm$7.7 & 96.3$\pm$3.2 & 228.0$\pm$10.9 & 4452$\pm$395 & 1566$\pm$377 \\ 
SN 2020abfe & 146 & 33.6$\pm$1.1 & 3.0$\pm$0.3 & 36.6$\pm$1.4 & 4411$\pm$389 & $<376$\\
SN 2020aekp & 132 & 149.5$\pm$4.0 & 33.0$\pm$1.0 & 182.5$\pm$5.0 & 7728$\pm$846 & $<833$\\
SN 2020aekp & 169 & 231.0$\pm$4.5 & 32.3$\pm$1.3 & 263.3$\pm$5.8 & 6775$\pm$839 & $<834$\\
SN 2020aekp & 211 & 251.0$\pm$9.5 & 58.6$\pm$3.4 & 309.6$\pm$12.8 & 7422$\pm$852 & 1342$\pm$836 \\ 
\hline
    \end{tabular}
    \label{tab:halpha}
\end{table*}

\begin{figure}
    \centering
    \includegraphics[width=0.48\textwidth]{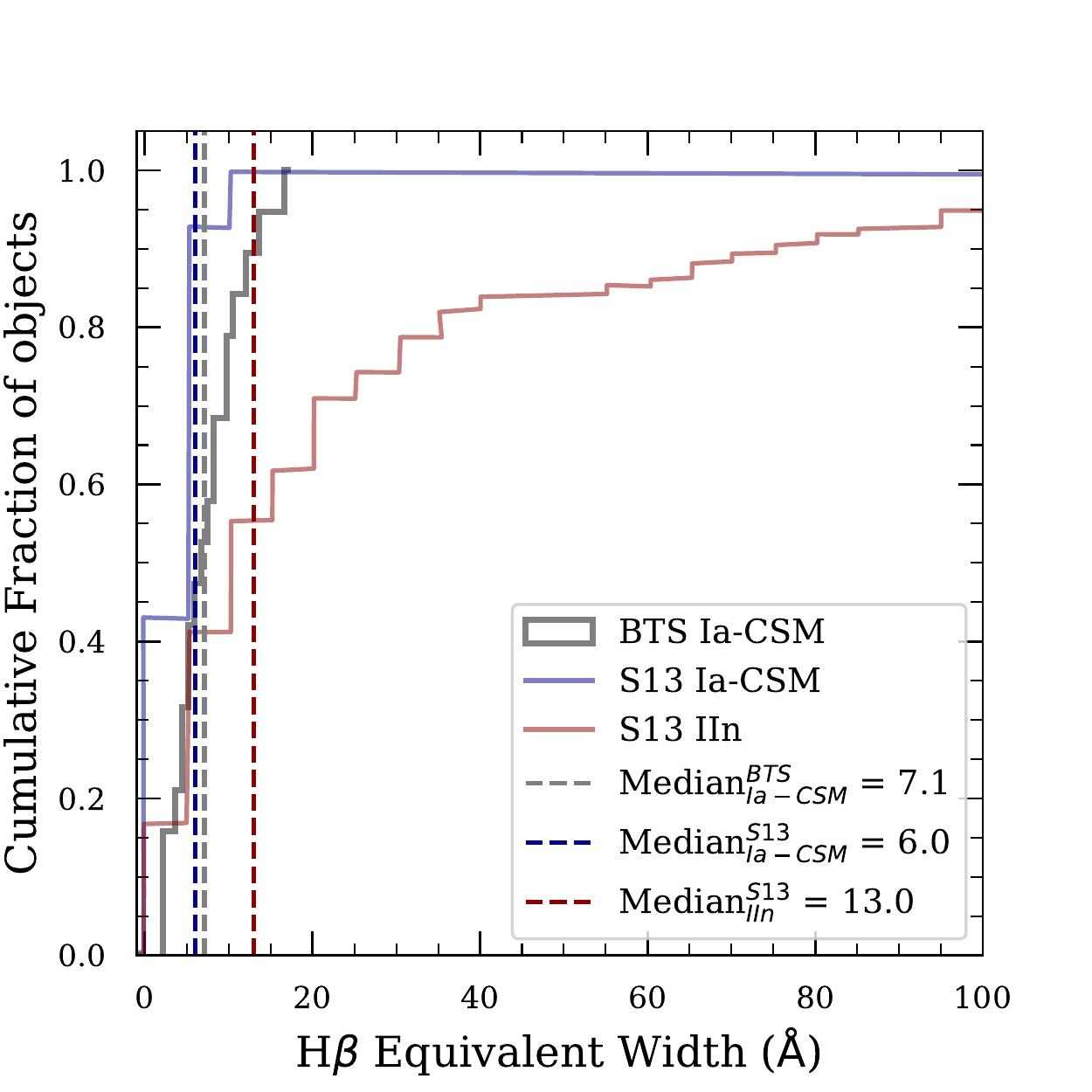}
    \includegraphics[width=0.48\textwidth]{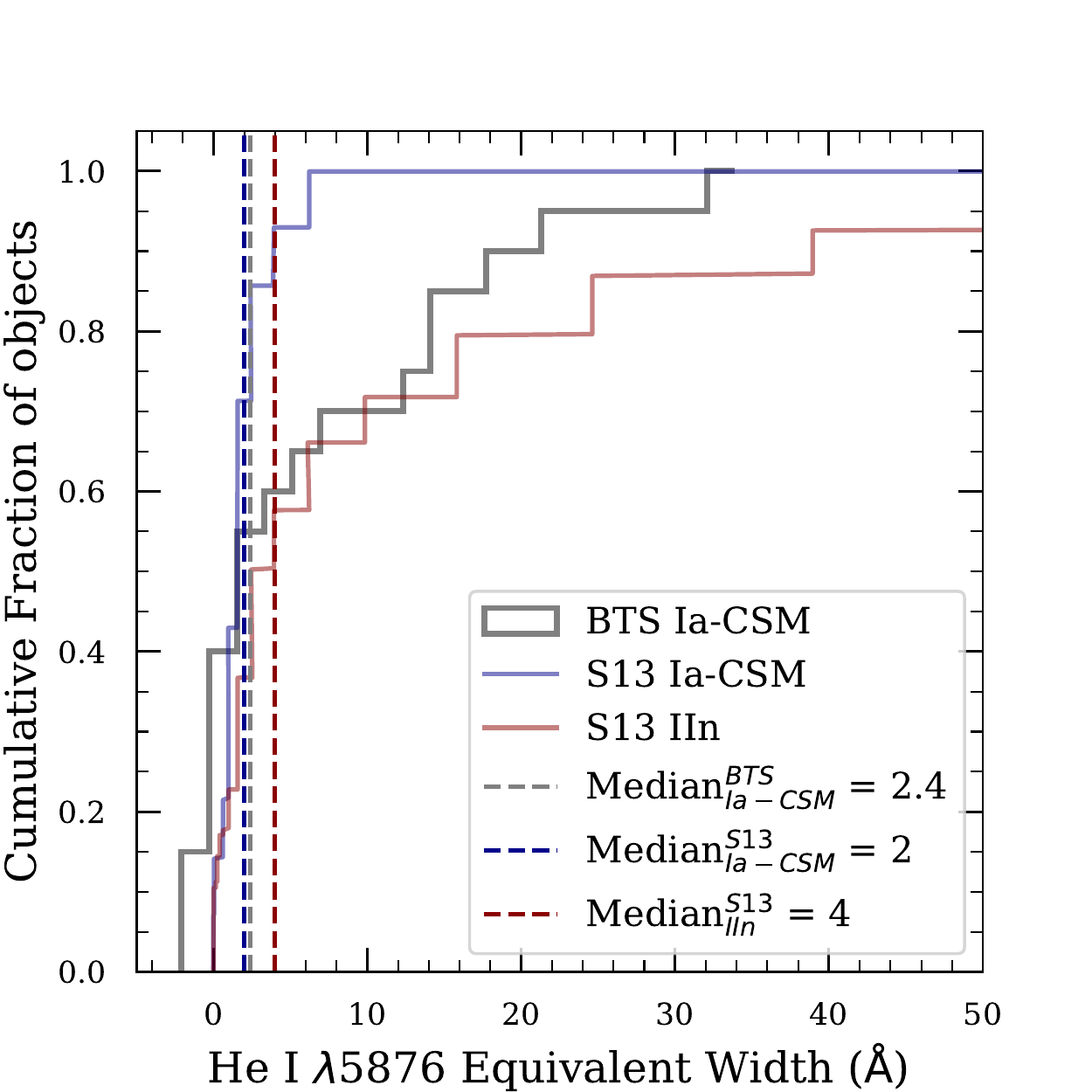}
    \caption{Cumulative distributions of equivalent width of H$\beta$ and \ion{He}{1} $\lambda$5876 emission lines calculated from the BTS SNe Ia-CSM (in grey) compared with the respective distributions presented in S13 for SNe Ia-CSM (blue) and SNe IIn (red). Vertical dashed lines mark the median EW of the distributions.}
    \label{fig:hbeta}
\end{figure}

The same analysis as for the H$\alpha$ line was also carried out for H$\beta$ and \ion{He}{1} $\lambda$5876 with a one component Gaussian fit. For cases where a Gaussian model could not fit the data, we integrate the flux value in a 100\,\AA\ region centered at 5876\,\AA\ for \ion{He}{1}. The \ion{Na}{1}D absorption lines are also prevalent in some spectra and blend with the \ion{He}{1} line, resulting in positive EWs for some SNe. The cumulative distributions of H$\beta$ and \ion{He}{1} equivalent widths are shown in the top and bottom panels of Figure~\ref{fig:hbeta} respectively.

The H$\beta$ median EW measured from the BTS SN Ia-CSM sample is 7.1\,\AA\,, close to the S13 value of $\sim$6\,\AA\ and quite weak compared to what S13 measured for SNe IIn ($\sim$13\,\AA\,). The overall cumulative distribution of H$\beta$ EW is also comparable to the S13 SNe Ia-CSM rather than to the S13 SNe IIn. For the \ion{He}{1} $\lambda$5876 line, the median EW measured for our BTS SN Ia-CSM sample, considering only significant emission features, is 2.4\,\AA\,. This is close to the value of $\sim$2\,\AA\ reported in S13, and again significantly different from their SN IIn value of $\sim$6\,\AA\, ($\sim$4\,\AA\, with upper limits), however the overall distribution seems to be closer to the S13 SNe IIn (but still weaker) rather than to the S13 SNe Ia-CSM. This indicates that perhaps \ion{He}{1} is not as good a discriminant between the populations compared to H$\beta$. Among the most He-rich SNe in our sample are SNe 2019ibk, 2020uem, 2020xtg, 2020aekp and 2018evt, and these SNe also have the higher H$\alpha$ equivalent widths in the sample. 

\begin{figure}
    \centering
    \includegraphics[width=0.48\textwidth]{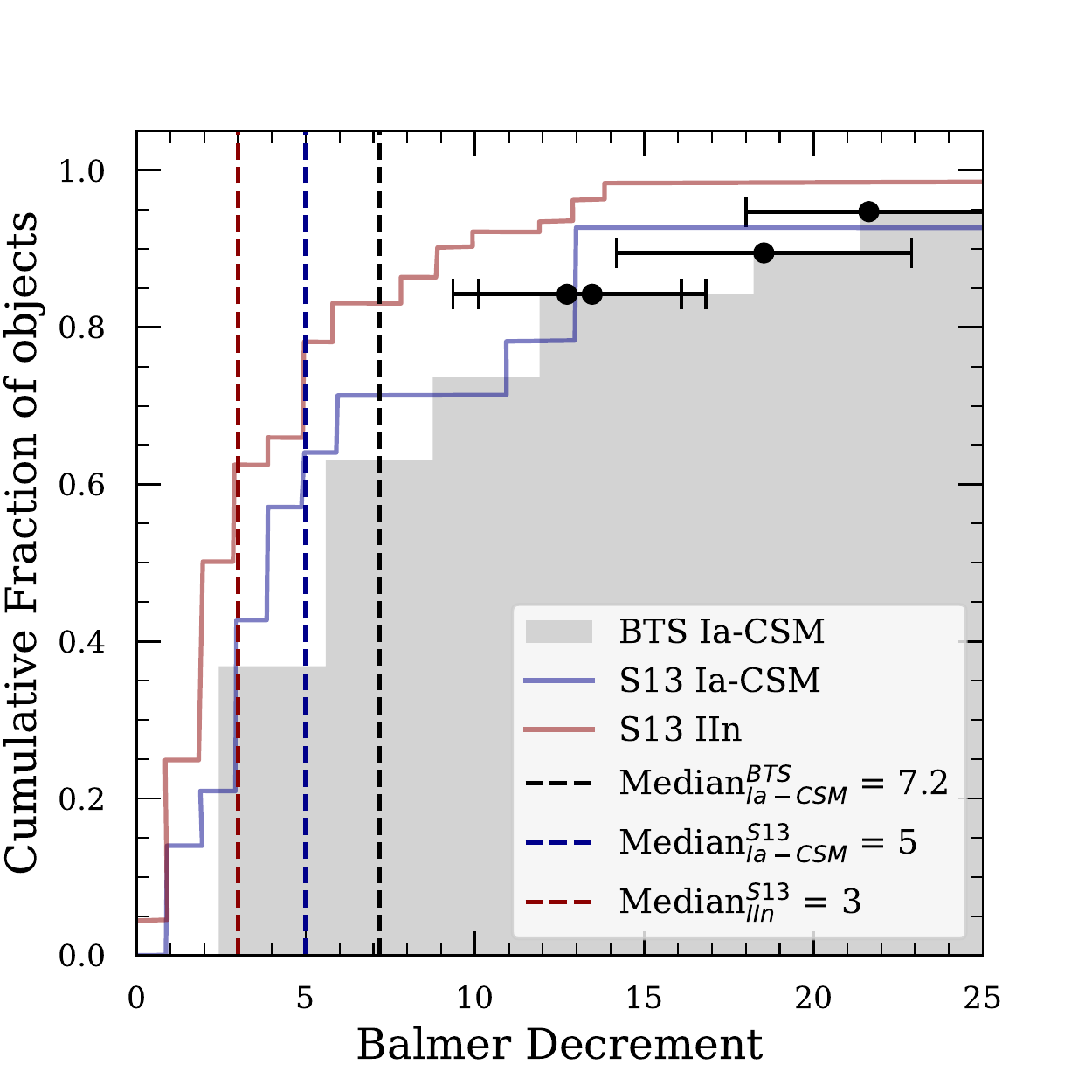}
    \caption{Cumulative distribution of $H\alpha / H\beta$ intensity ratio (Balmer decrement) calculated from intermediate resolution spectra of BTS SN Ia-CSM sample (grey shaded region). The red line is the distribution of Balmer decrement of SNe IIn measured in S13, the blue line is the SN Ia-CSM Balmer decrement distribution from S13. The black circles are a few representative points indicating the high Balmer decrement values and the uncertainties on them. The vertical dashed line is the median Balmer decrement measured from BTS SNe Ia-CSM.}
    \label{fig:badec}
\end{figure}

Figure~\ref{fig:badec} plots the cumulative distribution of the Balmer decrements ($\frac{F_{H\alpha}}{F_{H\beta}}$) measured for our sample SNe. The higher Balmer decrement values ($>$15) have large errors associated to them because of low SNR of the spectra from which they were derived, particularly near the H$\beta$ line. Consistent with the results of S13, the SNe Ia-CSM from this sample also have a high median Balmer decrement value of $\sim$7 ($\sim$5 in S13), indicating that the emission line mechanism is probably collisional excitation or self-absorption rather than recombination, from which the expected Balmer decrement value is $\sim$3. In the case of SNe Ia-CSM, if the CSM distribution consists of multiple shells as suggested for PTF11kx, moderately high densities could be created when fast moving ejecta overtake slowly moving thin dense CSM shells creating large enough optical depth in the H$\alpha$ line which results in the H$\beta$ transition decaying as Pa$\alpha$ $+$ H$\alpha$ \citep{Xu1992}. For some individual SNe where multiple spectra are available, the Balmer decrement is observed to first increase and later on decrease with phase. 

\subsection{Host galaxies}

We retrieved science-ready co-added images from the \textit{Galaxy Evolution Explorer} (GALEX) general release 6/7 \citep{Martin2005a}, the Sloan Digital Sky Survey DR 9 (SDSS; \citealt{Ahn2012a}), the Panoramic Survey Telescope and Rapid Response System (Pan-STARRS, PS1) DR1 \citep{Chambers2016a}, the Two Micron All Sky Survey \citep[2MASS;][]{Skrutskie2006a}, and preprocessed WISE images \citep{Wright2010a} from the unWISE archive \citep{Lang2014a}\footnote{\href{http://unwise.me}{http://unwise.me}}. 

We used the software package LAMBDAR (Lambda Adaptive Multi-Band Deblending Algorithm in R) \citep{Wright2016a} and tools presented in \citet{Schulze2021a}, to measure the brightness of the host galaxy. The spectral energy distribution (SED) was modelled with the software package Prospector\footnote{\href{https://github.com/bd-j/prospector}{\url{https://github.com/bd-j/prospector} version 0.3}} \citep{Johnson2021a}. We assumed a linear-exponential star-formation history, the \citet{Chabrier2003a} initial mass function, the \citet{Calzetti2000a} attenuation model, and the \citet{Byler2017a} model for the ionized gas contribution. The priors were set as described in \citet{Schulze2021a}.

\begin{figure}
    \centering
    \includegraphics[width=0.48\textwidth]{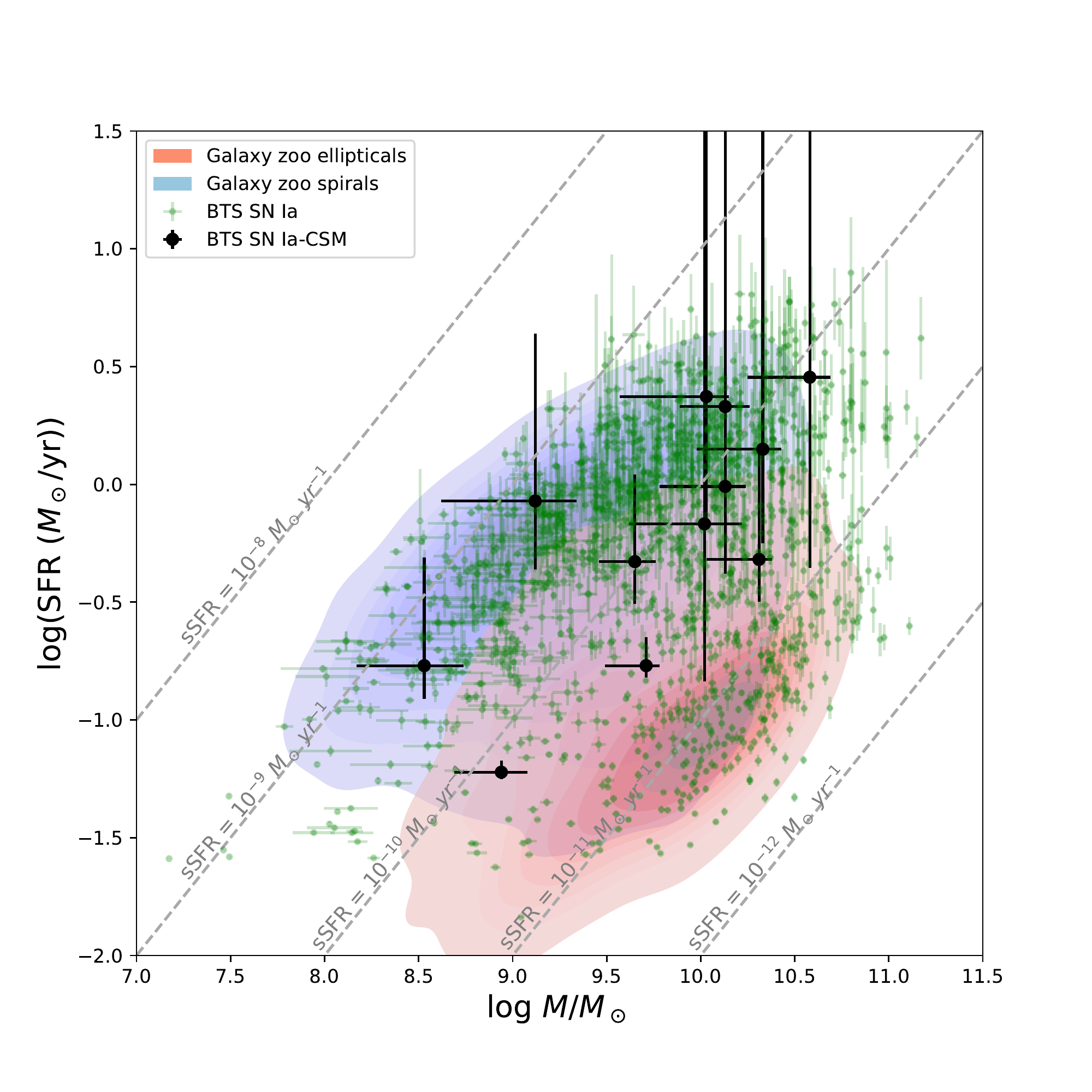}
    \caption{Host galaxies of BTS SN Ia-CSM (black circles) on SFR vs stellar mass plot with Galaxy-zoo spiral (blue contours) and elliptical (red contours) galaxies for comparison. BTS SN Ia hosts are also shown for comparison in green circles. Equal sSFR lines are marked with grey dashed lines.}
    \label{fig:hostgals}
\end{figure}

Figure \ref{fig:hostgals} shows the log of star formation rate (SFR) as a function of stellar mass for hosts of BTS SNe Ia-CSM. We also use a Galaxy-zoo \citep{galaxyzoo} sample of elliptical and spiral galaxies (randomly sampled in the redshift range $z=0.015-0.05$), and BTS SN Ia hosts as comparison samples collected by and used for comparison in \citet{irani2021}. We find the SN Ia-CSM host galaxy population to be consistent with late-type spirals and irregulars with recent star formation history. 4 out of 12 SNe have clearly spiral hosts, 3 have edge-on host galaxies, 4 seem to have irregulars as hosts and 1 has an unclear host type. Host galaxies of 10 out of 12 SNe have $w2-w3$ measurements available which are all $>1$ mag, putting them in late-type category \citep{irani2021}, 1 (SN 2019rvb) does not have W3 measurement but has $NUV-PS1_r\ \sim1$ mag again putting it towards late-type and 1 (SN 2020abfe) does not have any of the above information available except the $PS1_r$ band magnitude of 20.766, which is the faintest host galaxy (absolute SDSS $r$-band magnitude of $-17.4$) in our BTS SN Ia-CSM sample. As noted in S13, the SN Ia-CSM hosts of their sample had generally low luminosities ($-19.1 < M_r < -17.6$) except MW like spiral hosts. Our BTS SN Ia-CSM host luminosities lie in the range of $-21.8 < M_r < -17.4$ covering low to MW like luminosities.

\subsection{Rates}

Following the methodology for calculating the volumetric rate of transients found in the Bright Transient Survey from \citet{Perley2020}, we use their equation 2 to calculate the SN Ia-CSM rate:
$$ R = \frac{1}{T} \sum_{i=1}^N \frac{1}{(\frac{4\pi}{3}D^3_{max,i})f_{sky}f_{ext}f_{rec}f_{cl,i}} $$
where $T$ is the duration of the survey, $N$ is the number of transients that pass the quality cut, $D_{max,i}$ is the distance out to which the $i^{th}$ transient with peak absolute magnitude $M_i$ can be detected above the survey magnitude limit $m_{lim}$ (=19 mag for BTS SNe Ia-CSM) at peak light without any extinction, $f_{sky}$ is the average active survey coverage as a fraction of full sky, $f_{ext}$ is average reduction in effective survey volume due to Galactic extinction, $f_{rec}$ is the average recovery efficiency for a detectable transient within the survey coverage area, and $f_{cl,i}$ is the classification efficiency dependent on apparent magnitude.

The duration of the survey in which these 12 SNe Ia-CSM were detected is from 2018-05-01 to 2021-05-01, i.e. $T=3$ years. We calculate $f_{sky}$ during this time period by averaging the sky area coverage of the public MSIP survey considering 3 day cadence for ZTF Phase I (2018-05-01 to 2020-10-31) and 2 day cadence for ZTF Phase II (since 2020-11-01), which turns out to be 12505\,deg$^2$ for Phase I and 14831\,deg$^2$ for Phase II, giving a mean $f_{sky}=0.32$. We use the same value of 0.82 for $f_{ext}$ as calculated in \citet{Perley2020} given there has not been any change in the number and positions of ZTF fields.

To estimate $f_{rec}$, we consider SNe Ia-CSM brighter than $-18.5$ peak absolute magnitude and brighter than 18 apparent magnitude (total 5) of which 4 pass the quality cut, giving an $f_{rec}$ of 0.8. We take classification completeness of 0.75 at 19\,mag, 0.9 at 18.5\,mag and 1 at 17.2\,mag and linearly interpolate in between these values to get $f_{cl,i}$.

Then using $H_0=70$\,km\,s$^{-1}$\,Mpc$^{-1}$, ignoring cosmological effects\footnote{Contraction of control time window approximately compensated by increase in the star-formation rate density in the low redshift regime for redshift dependent SN rates.} as in \citet{Perley2020} and applying a uniform K-correction (K = 2.5$\times log_{10}(1+z)$), we get a rate of 29.35$^{+27.53}_{-21.37}$\,Gpc$^{-3}$\,yr$^{-1}$ for SNe Ia-CSM. We also calculate a SN Ia rate of 2.88$^{+0.28}_{-0.25}\times10^4$\,Gpc$^{-3}$\,yr$^{-1}$ from SNe Ia observed in the same period following the same method, which is close to the value of 2.35$\times10^4$\,Gpc$^{-3}$\,yr$^{-1}$ calculated in \citet{Perley2020}. This puts SNe Ia-CSM to be 0.02--0.2\% of SNe Ia. However this rate estimate should be considered a lower limit given various caveats in the correct identification of SNe Ia-CSM (see discussion \S\ref{sec:discuss3}). If the ambiguous classification cases outlined in Appendix~\ref{app:A} are considered to be SN Ia-CSM and included in the rate calculation, we obtain a rate upper limit of $97.7^{+135.8}_{-77.3}$\,Gpc$^{-3}$\,yr$^{-1}$, which is 0.07--0.8\% of SNe Ia.

\subsection{Precursor rates}
The ZTF precursor rates were calculated following the method in \citet{Strot2021} which studied the frequency of precursors in interacting SNe found in ZTF. \citet{Strot2021} included 6 of the SNe Ia-CSM presented in this paper in addition to 4 other SNe Ia-CSM not in this paper (see Appendix~\ref{app:A} for details) for their search but did not find any robust 5$\sigma$ precursor detections. This non-detection was concluded to be due to the small sample size of SNe Ia-CSM (or that they are more distant) compared to the SN IIn sample, so even if the precursors were as bright or frequent as for SNe IIn, it would be difficult to detect them. 

\begin{figure}
    \centering
    \includegraphics[width=0.48\textwidth]{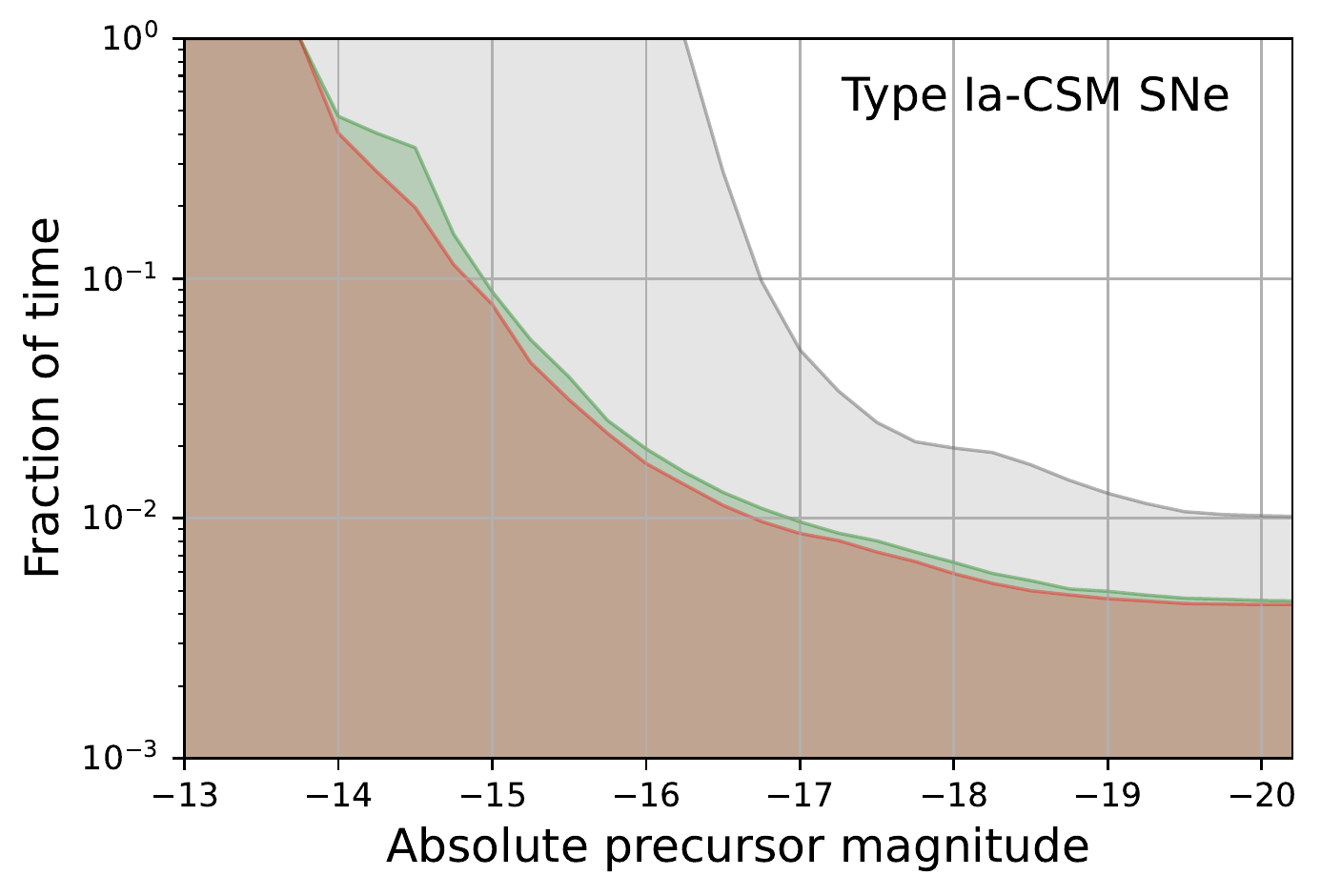}
    \caption{Precursor rate as a function of magnitude calculated from BTS SN Ia-CSM pre-explosion ZTF forced photometry stacked in 7-day bins. The different colored shaded regions correspond to different ZTF bands ($r$-red, $g$-green, $i$-grey). The solid lines depict the upper limits on fraction of the time a precursor of the corresponding magnitude would have been detected which is consistent with the ZTF non-detections.}
    \label{fig:precursor}
\end{figure}

The same search was here carried out for our larger sample by taking the ZTF forced photometry multi-band ($g,r,i$) light curves generated by the pipeline outlined in \citet{Masci2019} and stacking them in 1, 3 and 7-day long bins to search for faint outbursts. There were 7389 total available pre-explosion epochs for BTS SNe Ia-CSM, the earliest epoch being 1012 days prior to the explosion and the median phase 340 days prior. Hence the results are valid for typical SN Ia-CSM progenitors at about $\sim$1 year before the SN. We did not find any robust 5$\sigma$ precursor detections. The upper limits for the precursor rates in different bands are shown in Figure~\ref{fig:precursor}, where the solid lines indicate up to what fraction of the time a precursor of a given brightness could have been detected while being consistent with the ZTF non-detections. A precursor of $-15$ magnitude could occur as frequently as $\sim$10\% of the time given the ZTF non-detections. A continuous search for the precursors as more SNe Ia-CSM are found and classified and their sample size increases could yield a detection if the precursors are as frequent and bright as for SNe IIn. The dense and massive CSM around these objects is close enough to have been deposited within decades prior to the SN but the lack of precursors within 1 year indicates that there is likely no violent event that ejects a lot of mass in that period.
Probing for precursors could potentially constrain the progenitor in at least some cases. For example, \citet{Soker2013} predicts for their core degenerate (CD) model for PTF11kx-like SNe release of significant energy ($\sim$10$^{49}$\,erg) before explosion over timescale of several years, implying a precursor 3--7 magnitudes fainter than the SN explosion spread over several years, peaking in the near-IR.

\section{Discussion}\label{sec:discuss}
\subsection{Fraction of SNe Ia-CSM with delayed interaction}\label{sec:discuss1}
The fastest declining SNe in our sample (SNe 2018crl, 2020qxz and 2020aekp) are also the ones that develop a plateau and show relatively stronger SN Ia-like absorption features in their early spectra. They seem to have a delayed start for the interaction like PTF11kx but not as fast a decline, and thus bridge the gap between PTF11kx and the rest of the strongly interacting SNe Ia-CSM. It remains to be seen how many SNe Ia are weakly interacting where the CSM interaction starts in earnest at timescales of $\sim$year or more after explosion, this requires searching for faint detections in carefully calibrated forced photometry light curves (stacked to go fainter), a study currently undertaken by Terwel et al. (in prep). From the current sample, it appears that in addition to SNe Ia-CSM being intrinsically rare, delayed interaction SNe Ia-CSM are even rarer and only constitute about a quarter of all SNe Ia-CSM. This delayed interaction behaviour could also be an effect of asymmetric or clumpy CSM wherein part of the SN ejecta shine through depending on the viewing angle. Observational campaigns that capture the inner boundary of the CSM and the geometry robustly could shed light on the distribution of the inner CSM radius and reveal if it is a continuous distribution or if there are multiple progenitor scenarios within the SN Ia-CSM class.

\subsection{Implications for progenitor based on observed mass loss}\label{sec:discuss2}
From Figure~\ref{fig:massloss}, the estimated mass-loss rates from a simple spherical treatment of the CSM and a stationary wind lie between $\sim10^{-3}$ to $10^{-1}$\,M$_\odot$\,yr$^{-1}$ over a period of less than $\sim60$ years before explosion. That gives a total mass loss of $\sim0.1$ to $\sim1$\,M$_\odot$. \citet{dilday2012} estimated $\sim5$\,M$_\odot$ of CSM around PTF11kx while \citet{Graham2017} revised it to be $\sim0.06$\,M$_\odot$. Light curve modeling of SN 1997cy and SN 2002ic by \citet{ChugaiYungelson2004} resulted in $\sim5$\,M$_\odot$ estimates for both SNe. \citet{inserra2016} also fit analytical models to some SNe Ia-CSM and found the CSM mass to lie between 0.4 and 4.4\,M$_\odot$. Since from Figure~\ref{fig:bololum}, the pseudo-bolometric luminosities of our SNe Ia-CSM lie somewhere between PTF11kx and SNe 1997cy, 2002ic and 2005gj, with SN 1999E somewhere in the middle, we can say that the total CSM mass in our sample of SN Ia-CSM should also be several solar masses. A WD$+$AGB star system has typically been suggested for historical SNe Ia-CSM to explain this massive CSM. The WD could either gain mass through Roche Lobe overflow (RLOF) from the companion that drives an optically thick wind (OTW) or merge with the core of the AGB star that then explodes in or soon after the common envelope phase. \citet{MengPods2018} model WD$+$MS systems for their common envelope wind (CEW) model and find $\sim1$\,M$_\odot$ CSM around SNe Ia-CSM. Thus, given the large observed CSM mass range, the nature of the companion cannot be solely determined from total mass lost. High resolution spectroscopy that can resolve the narrow unshocked CSM wind velocity is also needed to determine the compactness of the companion.

\subsection{Implications for progenitor based observed volumetric rate}\label{sec:discuss3}
Robust observed rate estimates for SNe Ia-CSM have been few and far between. \citet{Dilday2010} found 1 interacting SN Ia (SN 2005gj) in a sample of 79 SNe Ia at $z < 0.15$ in the SDSS-II SN survey, giving a rate of $\sim$1\%. After the PTF11kx discovery in the Palomar Transient Factory (PTF) survey, the SN Ia-CSM rate was estimated to be $\sim$0.1\% (1 in 1000 classified SNe Ia; \citealp{dilday2012}) but without spectroscopic completeness determination. S13 identified 7 more SNe Ia-CSM from the PTF SN IIn sample, bumping up the estimate to $\sim$0.8\%. With this sample we have improved the rate estimate, providing a robust value (along with an uncertainty estimate on that value) from an unbiased survey with high spectroscopic completeness up to 18.5 magnitude. However this rate quite possibly still underestimates the true value for two reasons. The first being possible thermonuclear SNe that are enshrouded so completely by CSM interaction that they are misclassified as SNe IIn in the absence of good early time data. In the BTS SN IIn sample, we found 6 SNe IIn to have ambiguous classifications which could possibly be SNe Ia-CSM and these are described in Appendix~\ref{app:A}. Including these ambiguous cases in rate estimation results in a rate upper limit of 0.07--0.8\% for strongly interacting thermonuclear SNe, while excluding them gives an underestimated rate of 0.02-0.2\%.

The second issue with the rates is if there is indeed a continuum of delayed interaction SNe Ia-CSM like PTF11kx, interaction in SNe Ia may present itself hundreds of days later at magnitudes fainter than ZTF's limit ($\sim$20.5) resulting in those SNe not being counted when they may be sharing the same progenitor as the rest of the interacting SNe Ia-CSM. Lastly in some rare cases, the SN might appear normal in its light curve shape and duration (and thus would be missed by the selection criteria used in this paper) but seem to have peculiar narrow H$\alpha$ in its spectrum or bright mid-IR flux (like in the case of SN 2020aaym; \citealt{SN2020aaym_tns}). 

\citet{HanPods2006} predicted a rate of 0.1--1\% for 02ic-like events for their delayed dynamical instability SD model but could not naturally explain the delayed interaction and multiple CSM shells in PTF11kx (which is relevant for some SNe in our sample). A symbiotic nova-like progenitor was suggested by \citet{dilday2012} for PTF11kx and they quoted the theoretical rates for the same to lie between 1--30\%, however the model could not explain the massive CSM. \citet{Soker2013} suggested a core degenerate (CD) scenario in which the explosion is set by the violent prompt merger of the core of the giant companion on to the WD and could naturally explain the massive CSM of PTF11kx \citep{LivioRiess2003}. \citet{Soker2013} estimated the occurrence of such SNe (M$_{core} +$~M$_{WD} \gtrsim 2$\,M$_\odot$ and M$_{env} \gtrsim 4$\,M$_\odot$) through population synthesis and found it to be 0.002 per 1000\,M$_\odot$ stars formed. Assuming $\sim$1--2 SNe Ia occur per 1000\,M$_\odot$ stars formed \citep{MMB2012}, this corresponds to 0.1--0.2\%, which compares well with our observed rate estimate.

The CEW model by \citet{MengPods2018} predicts that the SNe Ia-CSM like objects could arise in the SD CEE scenario when CONe White Dwarfs (WD) steadily accrete material at the base of the CE without quickly spiraling in due to the driving of a CEW wind (10--100\,km\,s$^{-1}$). The WD explodes when it reaches the Chandrasekhar mass (1.38\,M$_\odot$) and could possibly explode within the CE before it is ejected. The CEW model predicts that 25--40\% of the SNe Ia from CONe WD in Common envelope evolution with a Main Sequence (MS) companion will show SN Ia-CSM like properties. \citet{MengPods2018} also give the ratio of SNe Ia from CONe WDs to normal SNe Ia from CO WDs to be between 1/9 and 1/5 (considering normal SNe Ia only come from CO WD + MS systems). Combining that with the estimate that roughly 10--20\% of all SNe Ia may come from the SD scenario \citep{Hayden2010,Bianco2011}, SNe Ia-CSM from CONe WD according to the CEW model should be 0.28\% to 1.6\% of all SNe Ia. A spin-down before explosion of the WD \citep{Justham2011,DistefanoKilic2012} could also explain the time delay between explosion and interaction.

\citet{Soker2022} estimated the common envelope to explosion delay time distribution (CEEDTD) shortly after the CEE (t$_{CEED}<10^4$\,yr) from SN in planetary nebula rates and SN Ia-CSM observed rates to be roughly constant rather than having a t$^{-1}$ dependence, that is the SN explosion could occur very soon after the CEE as well.  Our observed rates are on the lower side compared to these theoretical model estimates but compare well within the observational uncertainties, though the CEW model seems to best account for the overall SNe Ia-CSM properties.

\section{Summary}
In this paper, we have presented optical and mid-IR photometry, optical spectra and detailed analysis of 12 new SNe Ia-CSM identified in the Zwicky Transient Facility 
Bright Transient Survey, nearly doubling the total number of such objects discussed previously by \citet{Silverman_2013}. The properties of the sample extracted in this paper agree very well with similar analysis conducted in S13, particularly the median EW of H$\beta$ is found to be significantly weaker in SNe Ia-CSM compared with SNe IIn and consequently the Balmer decrements are ubiquitously higher in SNe Ia-CSM. The brightness of SNe Ia-CSM in mid-IR is comparable to SNe IIn and observations of reduced flux in the red side of the H$\alpha$ wing together with the mid-IR brightness points to formation of new dust in the cooling post-shock gas. The host galaxies of SNe Ia-CSM lie towards late-type galaxies with recent star formation. Unlike SNe IIn, no precursors were found within $\sim$1000 days before explosion for SNe Ia-CSM, which could be an observational bias (less number of SNe Ia-CSM compared to SNe IIn). We provide a robust rate estimate of 0.02--0.2\% of all SNe Ia for SNe Ia-CSM on account of the BTS survey being unbiased and spectroscopically highly complete. The simple mass-loss rate estimates from broad H$\alpha$ luminosity of $\sim10^{-2}$\,M$_\odot$\,yr$^{-1}$ are similar to previous estimates from various methods and indicate several solar masses of CSM around these SNe. The observed rate agrees well within the observational uncertainties with the CEW model by \citet{MengPods2018} which can also explain the interaction delay and massive CSM.

There are still many unanswered questions about the nature of the progenitors and if we are accurately identifying all potential members of this class. As ZTF Phase II continues, we are identifying more and more SNe Ia-CSM (interacting with hydrogen rich and helium rich CSM) and looking further to the future, if ZTF continues for a Phase III and when LSST survey operations begins, a larger sample would further improve upon the observed rate calculation. However, individual object studies are as important and detailed spectroscopic and multi-wavelength follow-up is essential to capture the CSM configuration and mass.

\section{Acknowledgment}
Based on observations obtained with the Samuel Oschin Telescope 48-inch and the 60-inch Telescope at the Palomar Observatory as part of the Zwicky Transient Facility project. ZTF is supported by the National Science Foundation under Grants No. AST-1440341 and AST-2034437 and a collaboration including current partners Caltech, IPAC, the Weizmann Institute of Science, the Oskar Klein Center at Stockholm University, the University of Maryland, Deutsches Elektronen-Synchrotron and Humboldt University, the TANGO Consortium of Taiwan, the University of Wisconsin at Milwaukee, Trinity College Dublin, Lawrence Livermore National Laboratories, IN2P3, University of Warwick, Ruhr University Bochum, Northwestern University and former partners the University of Washington, Los Alamos National Laboratories, and Lawrence Berkeley National Laboratories. Operations are conducted by COO, IPAC, and UW. The ZTF forced-photometry service was funded under the Heising-Simons Foundation grant \#12540303 (PI: Graham). 
 This work was supported by the GROWTH project (Kasliwal et al. 2019) funded by the National Science Foundation under PIRE Grant No 1545949. The Oskar Klein Centre was funded by the Swedish Research Council. Partially based on observations made with the Nordic Optical Telescope, operated by the Nordic Optical Telescope Scientific Association at the Observatorio del Roque de los Muchachos, La Palma, Spain, of the Instituto de Astrofisica de Canarias. Some of the data presented here were obtained with ALFOSC. Some of the data presented herein were obtained at the W. M. Keck Observatory, which is operated as a scientific partnership among the California Institute of Technology, the University of California, and NASA; the observatory was made possible by the generous financial support of the W. M. Keck Foundation. The SED Machine is based upon work supported by the National Science Foundation under Grant No. 1106171. This work has made use of data from the Asteroid Terrestrial-impact Last Alert System (ATLAS) project. The Asteroid Terrestrial-impact Last Alert System (ATLAS) project is primarily funded to search for near earth asteroids through NASA grants NN12AR55G, 80NSSC18K0284, and 80NSSC18K1575; byproducts of the NEO search include images and catalogs from the survey area. The ATLAS science products have been made possible through the contributions of the University of Hawaii Institute for Astronomy, the Queen’s University Belfast, the Space Telescope Science Institute, the South African Astronomical Observatory, and The Millennium Institute of Astrophysics (MAS), Chile. This research has made use of the NASA/IPAC Infrared Science Archive, which is funded by the National Aeronautics and Space Administration and operated by the California Institute of Technology. The Liverpool Telescope is operated on the island of La Palma by Liverpool John Moores University in the Spanish Observatorio del Roque de los Muchachos of the Instituto de Astrofisica de Canarias with financial support from the UK Science and Technology Facilities Council. 

Y. Sharma thanks the LSSTC Data Science Fellowship Program, which is funded by LSSTC, NSF Cybertraining Grant \#1829740, the Brinson Foundation, and the Moore Foundation; her participation in the program has benefited this work.

S. Schulze acknowledges support from the G.R.E.A.T research environment, funded by {\em Vetenskapsr\aa det},  the Swedish Research Council, project number 2016-06012.

This work has been supported by the research project grant “Understanding the Dynamic Universe” funded by the Knut and Alice Wallenberg Foundation under Dnr KAW 2018.0067,

The research of Y.\ Yang is supported through a Bengier-Winslow-Robertson Fellowship.

\texttt{Fritz} \citep{skyportal2019,duev2019real} and GROWTH marshal \citep{Kasliwal_2019} (dynamic collaborative platforms for time-domain astronomy) were used in this work.

\software{LAMBDAR \citep{Wright2016a}, Prospector \citep{Johnson2021a}, pySEDM \citep{pysedm2019}, IRAF \citep{Tody1986, Tody1993}, pyNOT (\url{https://github.com/jkrogager/PyNOT}), LPipe \citep{Perley2019}, pypeit \citep{pypeit:joss_pub}, extinction \citep{extinction2016}, pyraf-dbsp \citep{Bellm2016}, FPipe \citep{Fremling2016}, DBSP\_DRP \citep{dbsp_drp2021}, ztfquery \citep{ztfquery2018}, astropy \citep{astropy:2013, astropy:2018, astropy:2022}, matplotlib \citep{Hunter:2007}}.

\bibliographystyle{aasjournal}
\bibliography{biblio}

\appendix

\section{Ambiguous SN Ia-CSM/IIn in BTS}\label{app:A}
To identify potential SNe Ia-CSM hiding in the SN IIn sample classified by BTS, we rechecked all SNe IIn classifications (total 142) using SuperNova IDentification (SNID; \citealt{snid2007}) software. SNe IIn spectra were processed through SNID, and any SN having $\geq3$ matches to a SN Ia-CSM in the top 10 matches were manually checked. The SNe having ambiguous classifications are described below.
\subsection{SN 2019smj}
Discovered by ZTF and reported to TNS by ALeRCE \citep{alerce2021} on 2019-10-13 11:28:42.000, SN 2019smj (ZTF19aceqlxc) was classified as a Type IIn by BTS at $z=0.06$. It peaked at apparent magnitude $17.1$ in $r$ band ($\sim-20.1$) and then developed a weaker but broader bump. The spectra showed very weak H$\beta$, barely any \ion{He}{1} $\lambda5876$, no \ion{O}{1} $\lambda7774$ or [\ion{O}{1}] lines but showed some iron group lines, Ca NIR emission and [\ion{Ca}{2}]. SNID best matches were to SNe 1997cy and 2005gj. The early spectra from P60/SEDM have some matches to SN 2005gj but are too noisy and of ultra-low resolution to conclusively provide a Ia-CSM classification. From these observations, SN 2019smj is most likely a Type Ia-CSM but given the lack of confirmation we have excluded it from the main sample.

\subsection{SN 2018dfa}
Discovered and reported to TNS by ATLAS on 2018-07-05 08:51:21.000, SN 2018dfa was classified initially as a Type IIP by BTS but later spectra revealed it to be a Type IIn at $z=0.128$. It peaked at apparent magnitude of $17.5$ in $r$ band ($-20.2$) and showed a minor bump before main peak in the light curve. The spectra showed weak H$\beta$ and \ion{He}{1} $\lambda5876$, no \ion{O}{1} $\lambda7774$ or [\ion{O}{1}] lines. SNID best matches were to SNe 2002ic and 2005gj along with SNe Ia-norm/91T. The earliest spectra with good SNR from P200/DBSP had one match to SN 2005gj but could not provide a robust Ia-CSM classification. From these observations, SN 2018dfa is most likely a Type Ia-CSM but given the lack of confirmation we have excluded it from the main sample.

\subsection{SN 2019vpk}
Discovered by ZTF and reported to TNS by ALeRCE on 2019-11-25 06:33:38.000, SN 2019vpk was classified as a Type IIn by BTS at $z=0.1$. It peaked at apparent magnitude of $\sim18$ in $r$ band ($\sim-20.5$). The early spectra were too noisy and the only spectrum with good SNR was obtained with P200/DBSP nearly 6 weeks after discovery which showed weak H$\beta$, no clear \ion{He}{1} emission but possibly \ion{Si}{2} $\lambda5958$ emission (which is unlike any other SN Ia-CSM). SNID top matches were to SN 2005gj but visually did not look entirely convincing, and some matches were also to Type IIn. We conclude SN 2019vpk does not have enough data for a robust Ia-CSM classification.

\subsection{SN 2019wma}
Discovered by ZTF and reported to TNS by ALeRCE on 2019-12-13 13:35:26.000, SN 2019wma was classified as a Type IIn by BTS at $z=0.088$. It peaked at apparent magnitude of $\sim18.5$ in $r$ band ($\sim-19.5$). The spectra obtained were either from P60/SEDM or LT/SPRAT hence of low resolution and showed weak H$\beta$ and \ion{He}{1} emission. SNID top matches to earliest SEDM spectrum were to SN 2005gj at the correct redshift but given the lack of intermediate resolution spectra and absence of late time follow-up we did not assign a Type Ia-CSM classification to SN 2019wma and excluded it from the main sample.

\subsection{SN 2019kep}
Discovered and reported to TNS by ATLAS on 2019-07-02 14:13:55.000, SN 2019kep was classified as a Type IIn by BTS at $z=0.02388$. It peaked at apparent magnitude of $18.2$ in $r$ band ($-17$). Most early spectra were too noisy for classification but matched to SN 2005gj. A good SNR P200/DBSP spectrum showed narrow P-Cygni H$\alpha$ with absorption minimum at $\sim2500$\,km\,s$^{-1}$ but overall matched to a Type II SN. From these observations, we could not determine a robust classification for SN 2019kep and excluded it from the main sample.

\subsection{SN 2018ctj}
Discovered and reported to TNS by ZTF on 2018-04-21 08:36:57.000, SN 2018ctj was classified as a Type IIn by BTS at $z=0.0378$. It peaked at apparent magnitude of $18.4$ in $r$ band ($-17.8$) and was also detected in unWISE data. Only one P60/SEDM spectrum was obtained that matched well to SNe 1997cy and 2005gj. Given the lack of intermediate resolution spectra this SN remains classified as Type IIn and excluded from the main sample.

\end{document}